\documentclass[aps,pra,groupedaddress,showpacs,onecolumn,sort&compress,
floatfix, amssymb, noeprint]{revtex4-2}
\usepackage{float}
\usepackage[dvips]{graphicx}
\usepackage{dcolumn}
\usepackage{bm} 
\usepackage{color}
\usepackage{tabularx} 
\usepackage{booktabs} 
\usepackage[T1]{fontenc}
\usepackage[hidelinks]{hyperref}
\usepackage{upgreek}
\usepackage{xfrac}

\definecolor{gray}{rgb}{0.5, 0.5, 0.5}

\usepackage{bibunits}
\usepackage{caption}
 
\usepackage{subcaption}
\usepackage{graphicx}
\usepackage{amsmath}
\usepackage{amssymb}
\usepackage{multirow}
\usepackage{bm}
\usepackage[section]{placeins}
\usepackage{units}
\usepackage{svg}
\usepackage{xpatch}
\makeatletter
\xpatchcmd{\@ssect@ltx}{\@xsect}{\protected@edef\@currentlabelname{#8}\@xsect}{}{}
\xpatchcmd{\@sect@ltx}{\@xsect}{\protected@edef\@currentlabelname{#8}\@xsect}{}{}
\makeatother
\usepackage[normalem]{ulem}
\graphicspath{{./figures/}}
\DeclareMathOperator\Arg{Arg}

\begin{document}

\title{Accurate holographic light potentials using pixel crosstalk modelling}

\author{ P.\ Schroff, A.\ La Rooij \footnote{arthur.larooij@strath.ac.uk}, E.\ Haller, S.\ Kuhr\ }

\address{University of Strathclyde, Department of Physics, SUPA, Glasgow G4 0NG, United Kingdom}

\begin{abstract}
Arbitrary light potentials have proven to be a valuable and versatile tool in many quantum information and quantum simulation experiments with ultracold atoms. Using a phase-modulating spatial light modulator (SLM), we generate arbitrary light potentials holographically with measured efficiencies between $15\% -40\%$ and an accuracy of $<2\%$ root-mean-squared error. Key to the high accuracy is the modelling of pixel crosstalk of the SLM on a sub-pixel scale which is relevant especially for large light potentials. We employ conjugate gradient minimisation to calculate the SLM phase pattern for a given target light potential after measuring  the intensity and wavefront at the SLM. Further, we use camera feedback to reduce experimental errors, we remove optical vortices and investigate the difference between the angular spectrum method and the Fourier transform to simulate the propagation of light. Using a combination of all these techniques, we achieved more accurate and efficient light potentials compared to previous studies, and generated a series of potentials relevant for cold atom experiments.
\end{abstract}

\flushbottom

\maketitle

\thispagestyle{empty}

\section{Introduction}

The ability to shape light into arbitrary potentials has created many new opportunities in cold atom experiments.
Applications include atomtronics \cite{Kwon2020, Luick2020}, tailored potentials for quantum simulation experiments with optical lattices \cite{Guo2021, Sompet2022, Mazurenko2017, Navon2016} and quantum information platforms using Rydberg arrays \cite{Ebadi2021, Barredo2016, Amico2021}.
These applications require smooth light potentials that minimize inhomogeneities and the resulting dephasing effects, and for experiments involving larger atom numbers or where laser power is limited, a high efficiency is desirable. 
Arbitrary  light potentials are commonly generated using a digital micromirror device (DMD) which is an amplitude-modulating spatial light modulator (SLM) or using a phase-modulating liquid crystal on silicon (LCOS) SLM.
Tailored light potentials for cold atom experiments were realised using a DMD in direct imaging \cite{Gauthier2016, Gauthier2019}, where the efficiency of the light potential is directly proportional to the number of mirrors in the `on' position and is limited by the diffraction efficiency of the device (typically $30\%-88\%$) \cite{Hausler2017, Gauthier2016, Deng2022}. 
Alternatively, the DMD can be used in a holographic setup with efficiencies of $1\%-2\%$ \cite{Zupancic2016}. As opposed to direct imaging, any aberrations in the optical system can be corrected in situ which enables to generate diffraction-limited light potentials \cite{Zupancic2016}.
Using a phase-modulating LCOS SLM in a holographic setup, calculated efficiencies of $18\%-64\%$ were achieved \cite{Gaunt2012, Harte2014, DeMarco2008, Bowman2017}, largely independent of the size of the light potential. After multiplying these calculated efficiencies by the diffraction efficiency of the LCOS SLM ($20\%-90\%$, depending on the diffraction angle \cite{Ronzitti2012}) they are still an order of magnitude higher compared to the DMD efficiencies.
As holographically generated light potentials are very sensitive to aberrations in the optical system, it is challenging to produce light potentials of low error. Potentials with a root-mean-squared (RMS) error of <5\% have been used to investigate Bose-Einstein condensates in ring traps  \cite{Wright2013, Bruce2011}, while in recent experiments with Rydberg arrays, light potentials with an RMS error of 0.7\% were used \cite{Ebadi2021}. Despite the complexity associated with a holographic setup, the prospect of achieving higher efficiencies and lower error has driven the development of sophisticated hologram calculation techniques.

The task of finding the SLM phase to achieve a desired light potential is known as phase retrieval problem. Various algorithms such as the mixed-region amplitude-freedom (MRAF) algorithm \cite{DeMarco2008}, the offset-MRAF algorithm \cite{Gaunt2012} and a conjugate gradient (CG) approach \cite{Harte2014} were developed to solve this purely computational problem and produce simulated light potentials of $<1\%$ RMS error. However, creating light potentials with this degree of accuracy is difficult experimentally as imperfections in the optical setup cause a mismatch between the simulated and the measured light potentials. These effects  include a distorted wavefront at the SLM, the curved surface of the SLM itself, crosstalk between neighbouring SLM pixels, aberrations caused by the Fourier lens and other alignment imperfections. To compensate for these errors, camera feedback algorithms were used to create more accurate light potentials \cite{Bruce2011, Bruce2015, Bowman2018, Ebadi2021, DeMarco2008}. Using stochastic gradient descent, the phase retrieval problem was solved by directly taking the camera image into account when calculating the cost function and its gradient \cite{Peng2020}. Further, it was shown that the Fourier transform used to propagate the light field from the SLM to the camera can be replaced by a more sophisticated method to simulate the propagation of light and can result in more accurate experimental light potentials \cite{Gaunt2012}. 

In this work, we create light potentials by combining several computational and experimental techniques to achieve an RMS error of $<2\%$ for various patterns while maintaining measured efficiencies between $15\% - 40\%$. 
We solve the phase retrieval problem by using CG minimisation \cite{Harte2014, Bowman2017} and investigate the difference between two methods to simulate the propagation of the light; the angular spectrum method (ASM) \cite{Goodman2017} and the commonly used fast Fourier transform (FFT). We further improve the quality of our light potentials by modelling crosstalk of neighbouring SLM pixels. In previous work, spot patterns have been generated using a pixel crosstalk model \cite{Ronzitti2012}, however, to the best of our knowledge, this effect has not been taken into account to generate smooth arbitrary light potentials.
The combination of all of these techniques allows us to produce potentials of  $<1.5\%$ RMS error and efficiencies of more than $40\%$, opening the way to new applications that require this degree of accuracy and efficiency.
 
\begin{figure}[!tb]    \centering{\phantomsubcaption\label{fig:phase_retrieval}\phantomsubcaption\label{fig:flow_diagram}\phantomsubcaption\label{fig:convergence}}
    \includegraphics[width=12cm]{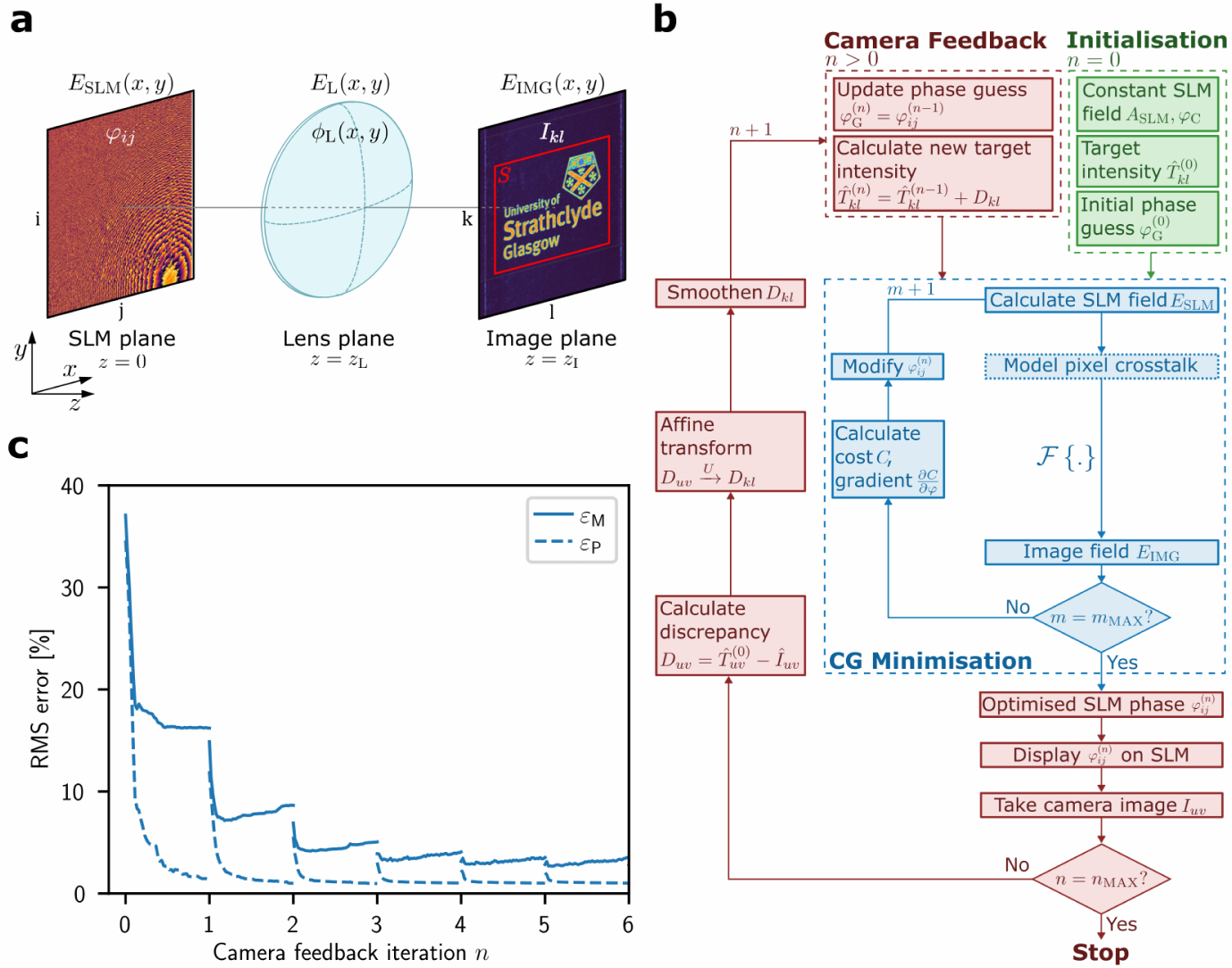}
    \caption{Generation of light potentials using CG minimisation and camera feedback. (\subref{fig:phase_retrieval}) Holographic setup with the displayed SLM phase, $\varphi_{ij}$, in the SLM plane, forming a light potential, $I_{kl}$, in the Fourier plane of the lens. (\subref{fig:flow_diagram}) Flow diagram visualising the process of generating a light potential. The pixel crosstalk on the SLM is modelled just before the SLM field, $E_{\text{SLM}}$, is propagated to the image plane. (\subref{fig:convergence}) Convergence of the first $6$ iterations of the feedback process. Due to imperfections, the experimental RMS error, $\varepsilon_{\text{M}}$, (solid line) converges at a higher level than the predicted RMS error, $\varepsilon_{\text{P}}$ (dashed line). After each feedback iteration, $n$, the experimental RMS error, $\varepsilon_{\text{M}}$, decreases due to the adjusted target light potential.}
\label{fig:algorithm}
\end{figure}

Our experimental setup consists of the SLM (Hamamatsu X13138-07, pixel pitch $\unit[12.5]{\mu m}$, $1272 \times 1024$ pixels), an achromatic doublet lens and a camera in the Fourier plane. A full description of the setup is shown in the SI. The electric field in the SLM plane, $E_{\text{SLM}}\!\left(x, y\right)$, is related to the electric field in the image plane, $E_{\text{IMG}}\!\left(x, y\right)$, via the Fourier transform (details see Fig. \ref{fig:algorithm} and \nameref{sec:methods}). To generate the desired light potential with an intensity pattern $I_{\text{IMG}}\!\left(x, y\right)=\left| E_{\text{IMG}}\!\left(x, y\right)\right|^2$, the phase pattern displayed by the SLM, $\varphi\!\left(x, y\right)$, must be found, given the constant field at the SLM $A_{\text{SLM}}\!\left(x, y\right)\exp{\left[ i\varphi_{\text{C}}\!\left(x, y\right)\right]}$. The Gerchberg-Saxton (GS) algorithm \cite{Gerchberg1972} is an iterative Fourier transform algorithm (IFTA) and can find $\varphi\!\left(x, y\right)$ to produce spot patterns of $98\%$ uniformity \cite{Kim2019}. However, for arbitrary and smooth light potentials, required e.g., for quantum simulation experiments with ultracold atoms, the GS algorithm does not converge well. Modified versions of the original GS algorithm such as the mixed-region amplitude-freedom (MRAF) algorithm \cite{DeMarco2008} and the offset-MRAF (OMRAF) algorithm \cite{Gaunt2012}, have produced smooth simulated light potentials approaching 1\% root-mean-square (RMS) error, and predicted efficiencies around 24\% \cite{Gaunt2012}, depending on the target pattern (see equation \ref{eq:rmse} and \ref{eq:eff}). More recently, gradient-based optimisation algorithms such as the CG method were used to generate simulated light potentials $<0.1\%$ RMS error and efficiencies  $>60\%$ \cite{Harte2014}, outperforming the above-mentioned IFTAs \cite{DeMarco2008, Gaunt2012}. Note that these are RMS errors and efficiencies of simulated light potentials which differ from the experimentally obtained values (see Table \ref{tab:lit}).

\section{Results}
\label{sec:results}

\subsection{Characterisation of light potentials}
\label{sec:characterisation}

To characterise the quality of our light potentials, we define the predicted and measured RMS error, $\varepsilon_{\text{P}}$ and $\varepsilon_{\text{M}}$, respectively,
\begin{equation}
\varepsilon_{\text{P}} = \sqrt{\frac{1}{N_M}\sum_{k,l\in M}\frac{\left(\hat{T}_{kl}-\hat{I}_{kl}\right)^2}{\hat{T}_{kl}^2}} \text{ \;  \; and  \;   \; }  
\varepsilon_{\text{M}} = \sqrt{\frac{1}{N_{U}}\sum_{u,v\in M_{U}}\frac{\left(\hat{T}_{uv}-\hat{I}_{uv}\right)^2}{\hat{T}_{uv}^2}}.
\label{eq:rmse}
\end{equation}
The predicted RMS error, $\varepsilon_{\text{P}}$, measures the difference between the simulated light potential, $\hat{I}_{kl}$, and the target potential, $\hat{T}_{kl}$, where $k$ and $l$ are the indices in the computational image plane. The error is evaluated in a measure region, $M$, which is defined as the region in the image plane where the target intensity is larger than 50\% of the maximum target intensity \cite{Bruce2015}. The number of pixels in $M$ is indicated by $N_M$. $\hat{T}_{kl}=T_{kl} / \sum_{k,l \in M} T_{kl}$ and $\hat{I}_{kl}=I_{kl} / \sum_{k,l \in M} I_{kl}$ are the normalised target light potential and the normalised simulated light potential, respectively.
The measured RMS error, $\varepsilon_{\text{M}}$, characterises the light potential captured by the camera. The camera image, $I_{uv}$, with row and column indices, $u$ and $v$, is mapped to the computational image plane using an affine transform, $U$. We define $\varepsilon_{\text{M}}$ in the transformed measure region, $M_{U}$, containing $N_{U}$ pixels, using the normalised, transformed target light potential, $\hat{T}_{uv}$.

The predicted efficiency, $\eta_{\text{P}}$, of the light potential is given by the ratio of the power in the signal region, $S$, (indicated by the red rectangle in Fig.\,\ref{fig:phase_retrieval}), to the total power in the image plane \cite{DeMarco2008}. 
We define the experimental efficiency of the light potential, $\eta_{\text{M}}$, as the ratio of optical power, $P_S$, in the transformed signal region, $S_U$, to the measured power of the beam before the expansion telescope, $P_{\text{in}}$, (see \nameref{sec:methods})
\begin{equation}
\eta_{\text{P}} = \frac{\sum_{k,l\in S}I_{kl}}{\sum_{k,l}I_{kl}}   \text{ \;  \; and  \;   \; } 
\eta_{\text{M}} = \frac{P_S}{P_{\text{in}}}.
\label{eq:eff}
\end{equation}

\subsection{Conjugate gradient minimisation and camera feedback}
\label{sec:cg_fb}

We use CG minimisation \cite{Harte2014} due to its rapid convergence and due to its flexibility to define a cost function which can be chosen to meet the requirements for a specific application, e.g., the optimisation of intensity, phase and efficiency in a specific region of interest. The minimisation improves the simulated light potential, $I_{kl}$, iteratively by modifying the SLM phase, $\varphi_{ij}$, based on a cost function $C$ and its gradient $\partial C / \partial \varphi_{ij}$ (blue loop in Fig. \ref{fig:flow_diagram}). We use the mean-squared error between the normalised simulated intensity pattern in the image plane, $\tilde{I}_{kl}=I_{kl} / \sum_{k,l \in S} I_{kl}$, and the normalised target intensity pattern, $\tilde{T}_{kl}=T_{kl} / \sum_{k,l \in S} T_{kl}$, in the signal region, $S$, as cost function for the optimisation \cite{Harte2014},
\begin{equation}
C\left(\varphi\right) = s \sum_{k,l\in S} \left( \tilde{T}_{kl}-\tilde{I}_{kl} \right)^2.
\label{eq:cost}
\end{equation}
The sum is evaluated over $k$ and $l$ in the signal region, $S$, where $s$ is the steepness of the cost function to aid convergence (see \nameref{sec:methods} for further details). 

To generate light potentials of low RMS error experimentally, it is essential to measure the beam profile, $A_{\text{SLM}}\!\left(x, y\right)$, and constant phase, $\varphi_{\text{C}}\!\left(x, y\right)$, at the SLM plane. We use an interferometric method \cite{Zupancic2016} which displays a sequence of patterns on subsections on the SLM (see SI). Finding a suitable initial SLM phase guess is essential for the convergence of the CG minimisation.
We choose an initial phase guess for a given light potential and remove optical vortices from a light potential if necessary (see \nameref{sec:methods}). To reduce the error in the experimental light potential further, we employ a camera feedback algorithm \cite{Bruce2015} (red loop in Fig. \ref{fig:flow_diagram}, details see \nameref{sec:methods}). The entire protocol is shown schematically in Fig. \ref{fig:flow_diagram}. After $m_{\text{max}}$ CG iterations, a camera image is taken to update the target image and restart the CG loop. The feedback algorithm typically converges within $n=15$ iterations (see Fig. \ref{fig:convergence}).

\subsection{Pixel crosstalk modelling}
\label{sec:crosstalk}

\captionsetup[figure]{aboveskip=-5pt}
\begin{figure}[!t]
    \centering{\phantomsubcaption\label{fig:fft_pred}\phantomsubcaption\label{fig:ct_cam}\phantomsubcaption\label{fig:ct_pred}}
    \includegraphics[width=14cm]{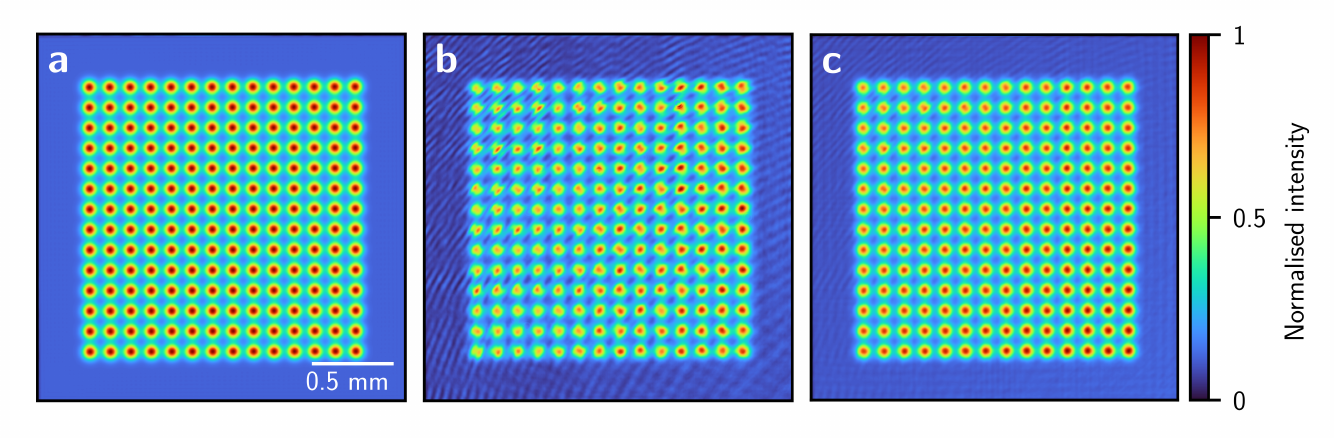}
    \caption{Simulated and experimental images illustrating the effect of pixel crosstalk. (\subref{fig:fft_pred}) Simulated light potential for a spot array target light potential. (\subref{fig:ct_cam}) Camera image of the experimental light potential showing fringes and an intensity gradient, with less intense spots in the top left of the image. (\subref{fig:ct_pred}) Simulated light potential after up-scaling and convolving the SLM phase pattern with kernel $K$. The fringes and the intensity gradient seen in the camera image (\subref{fig:ct_cam}) are reproduced in the simulation, however, with reduced contrast.}
    \label{fig:ct}
\end{figure}

\captionsetup[figure]{aboveskip=-5pt}
\begin{figure}[!b]
    \centering{\phantomsubcaption\label{fig:disc_small}\phantomsubcaption\label{fig:disc_medium}\phantomsubcaption\label{fig:disc_large}\phantomsubcaption\label{fig:disc_rms}\phantomsubcaption\label{fig:disc_profile}}
    \includegraphics[width=14cm]{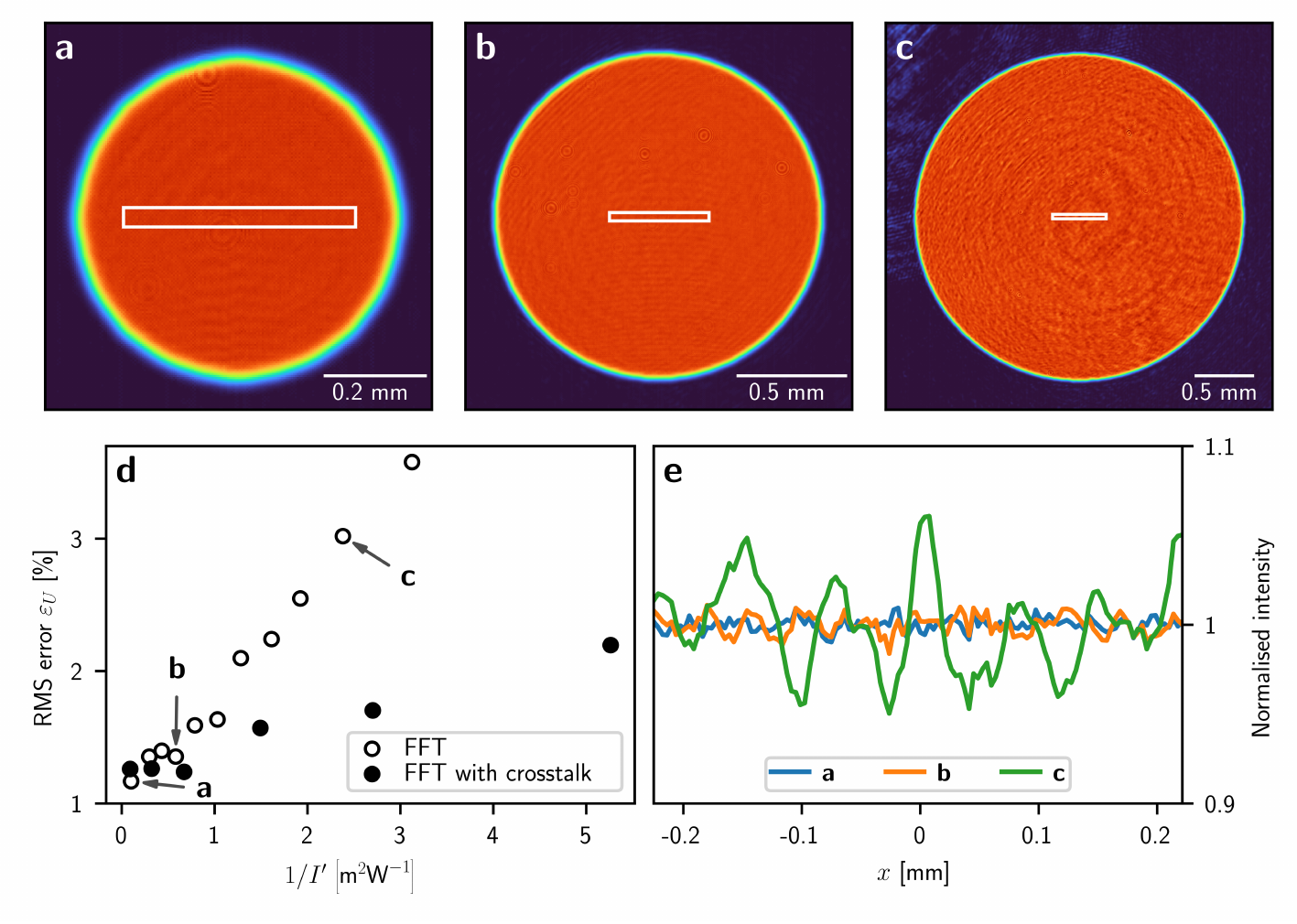}
    \caption{Effect of pattern size and pixel crosstalk on the RMS error. (\subref{fig:disc_small})-(\subref{fig:disc_large}) Disc-shaped potentials (diameters $D=0.6\,\text{mm}$, $D=1.5\,\text{mm}$ and $D=2.8\,\text{mm}$), generated using camera feedback without the pixel crosstalk model, and normalised by the average intensity in the flat part of the disc. (\subref{fig:disc_rms}) RMS error of disc-shaped light potentials of different diameters with and without the pixel crosstalk model. (\subref{fig:disc_profile}) Horizontal profiles of the light potentials, averaged over 10 rows within the white rectangles in (\subref{fig:disc_small})-(\subref{fig:disc_large}).}
    \label{fig:disc}
\end{figure}

By modelling a single SLM pixel with a single computational pixel, we assume that the phase across the SLM pixel is uniform.
However, due to the nature of the liquid crystal material inside the SLM, neighbouring pixels affect each other at their boundary region.
This effect is known as pixel crosstalk or fringing field effect \cite{Bengtsson2007, Persson2012, Moser2019, Pushkina2020, Guesmi2021, Moreno2021}.
We study the effect of pixel crosstalk on our light potentials by  up-scaling the SLM phase such that one SLM pixel is represented by $3 \times 3$ computational pixels and convolving it with a kernel, $K$, \cite{Moreno2021}
\begin{equation}
K\left(x, y\right) = \mathcal{F}^{-1}\left\{\exp\left[-\left(\frac{\left|\kappa_x\right|^q+\left|\kappa_y\right|^q}{\sigma^q}\right)\right]\right\},
\label{eq:ct_kernel}
\end{equation}
of order $q$ and width $\sigma$. As an example, we calculated the SLM phase for a spot array target potential using the CG minimisation, and observed fringes in the camera image (Fig. \ref{fig:ct_cam}) which do not appear in the simulated light potential (Fig. \ref{fig:fft_pred}).
After up-scaling and convolving the same SLM phase pattern, we propagate the field from the SLM plane to the image plane using the Fourier transform. Modelling the pixel crosstalk has no influence on the spatial resolution of the light potential in the image plane. 
The resulting simulated light potential (Fig. \ref{fig:ct_pred}) features fringes similar to those in the camera image, however, with reduced contrast.

In the CG minimisation, we account for pixel crosstalk by upscaling the displayed phase, $\varphi\left(x, y\right)$, and restricting its values to a range between $0$ and $2\pi$ to ensure that the cost $C\left(\varphi\right)$ remains a continuous, differentiable function.
We convolve the up-scaled phase with the kernel, $K$, before propagating the light field to the image plane.
The parameters $\sigma=1.24 \,\text{px}^{-1}$ and $q=1.80$ were found by a 2D scan to minimise $\varepsilon_{\text{M}}$ after 150 CG iterations without camera feedback for a disc-shaped target potential.
Using the camera feedback algorithm with the pixel crosstalk model further reduces the RMS error. The final RMS error and the effect of modelling the pixel crosstalk depend on the size of a specific target light potential (see Fig. \ref{fig:disc}).
Upscaling the SLM pixels by a factor of $3$ is computationally expensive, however, we accelerate our calculations using a GPU. This reduces the runtime of our algortihm to $\sim$ 10 minutes which is 6 times longer than without pixel crosstalk modelling (for 15 feedback iterations with 100 CG iterations each).

\captionsetup[figure]{aboveskip=3pt}
\begin{figure}[!b]
    \centering
    \includegraphics[width=14cm]{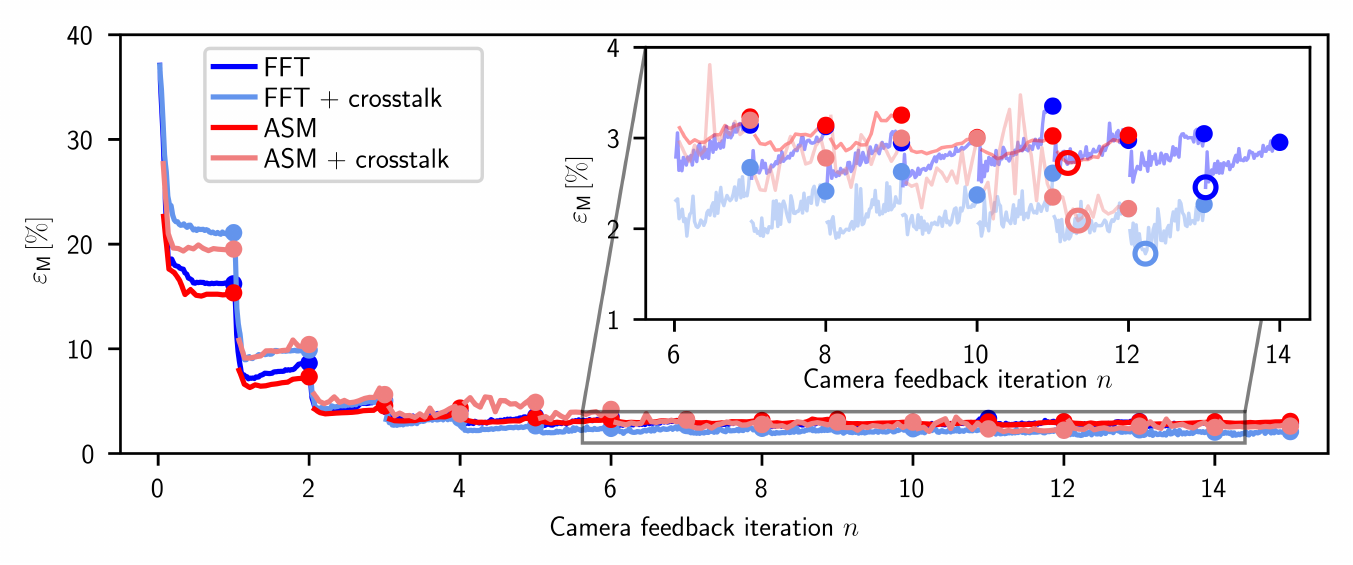}
    \caption{
    Convergence of the feedback procedure using the FFT and the ASM, with and without pixel crosstalk modelling. The main figure shows $\varepsilon_{\text{M}}$ as it converges for $n=15$ camera iterations with $m=100$ CG iterations in between.
    The values $\varepsilon_{\text{M}}$ used in the camera feedback process are shown as filled circles. To investigate the behaviour of $\varepsilon_{\text{M}}$ during the CG minimisation, we saved intermediate phase patterns and analysed the resulting light potentials (lines in main figure). For $n>1$, the experimental error, $\varepsilon_{\text{M}}$, is smallest for $m < 100$. The inset shows the convergence during the final 8 camera feedback iterations. The lowest experimental error was found between $n=11$ and $n=14$ (hollow circles in the inset).
}
    \label{fig:fft_asm_ct}
\end{figure}

To study how the pixel crosstalk model affects our light potentials, we produced disc-shaped light potentials of different diameters, $D$, between $\unit[0.64]{mm}$ and $\unit[3.3]{mm}$, with and without accounting for pixel crosstalk (see Fig. \ref{fig:disc}). The target light potential was convolved with a Gaussian kernel of 2 pixels width to ensure that the edge of the disc is not sharper than the diffraction limit. For the initial phase guess, the quadratic phase curvature was adjusted proportional to the disc diameter (see \nameref{sec:methods}). This ensures that the predicted efficiency of the differently sized light potentials remain similar ($\eta_{\text{P}} = 74\% -87\%$). Without accounting for pixel crosstalk, we achieved best light potentials ($\varepsilon_{\text{M}}=1.1\%$) for small discs of $D=\unit[0.85]{mm}$, and less smooth potentials ($\varepsilon_{\text{M}}=3.6\%$) for larger discs of $D=\unit[3.2]{mm}$  with measured efficiencies $\eta_{\text{M}}=33\%-40\%$. We found that $\varepsilon_{\text{M}}$ is inversely proportional to the measured intensity, $I'$, in the flat part of the disk (Fig. \ref{fig:disc_rms}). To obtain $I'$, we measure the average intensity in the flat part of the disc using the camera (see \nameref{sec:methods}). Smaller discs are of higher intensity since the same amount of optical power is focussed onto a smaller area. 

We found that the pixel crosstalk causes a ghost image \cite{VanBijnen2013, Ronzitti2012} which can interfere with the light potential and cause fringes. By accounting for pixel crosstalk in our model, any interference with the light potential caused by the ghost image is attenuated which lowered the final experimental RMS error by a factor of $\sim 0.4$ ($D=\unit[2.8]{mm}$). We found that accounting for pixel crosstalk has little effect on smaller light potentials, where the overlap between the ghost image and the light potential is smaller (see Fig. \ref{fig:disc_rms}). When taking the pixel crosstalk model into account, the RMS error, $\varepsilon_{\text{M}}$, remains smaller as the ghost image caused by pixel crosstalk is attenuated (Fig. \ref{fig:disc_rms}), and the measured efficiency decreases from $\eta_{\text{M}}=41\%$ ($D=\unit[0.85]{mm}$) to $\eta_{\text{M}}=20\%$ ($D=\unit[3.2]{mm} $). We found that the predicted efficiency, $\eta_\text{P}$, is proportional to the measured efficiency, $\eta_{\text{M}}$. The efficiency predicted by the pixel crosstalk model is lower and closer to the measured efficiency as multiple diffraction orders are simulated. We did not see an improvement in $\varepsilon_{\text{M}}$ when increasing the resolution of an SLM pixel even further to $5\times 5$ or $7\times 7$ computational pixels.

\subsection{Angular spectrum method}
\label{sec:asm}

The Fourier transform used to compute the propagation of light from the SLM plane to the image plane requires the far-field and the paraxial approximations (including a parabolic lens) as well as the assumption that lens and camera are perfectly in focus. In practice, we use a doublet lens and and there is an experimental position uncertainty of the lens and the camera along the optical axis. Inspired by the improvement in RMS error reported in a recent study \cite{Gaunt2012}, we implement Helmholtz propagation using the angular spectrum method (ASM) to model the diffraction of light without assuming a far field or small angles \cite{Goodman2017} (see SI for details on the ASM). 
In our method, this replaces the Fourier transform, $\mathcal{F}$, in the CG minimisation (shown in blue in Fig. \ref{fig:flow_diagram}) with the ASM.

We investigate the effect of using the ASM together with the pixel crosstalk model in our feedback process (Fig. \ref{fig:fft_asm_ct}). The ASM is more accurate than the FFT before any camera feedback is used ($n=0$ in Fig. \ref{fig:fft_asm_ct}), however, both methods converge to similar values after 15 iterations (see inset in Fig. \ref{fig:fft_asm_ct}). When including the pixel crosstalk model, the initial error before camera feedback ($n=0$) is higher, but the algorithm converges to lower $\varepsilon_{\text{M}}$ after 15 iterations for both the ASM and the FFT method. In all methods, the experimental error, $\varepsilon_{\text{M}}$, slowly rises towards the end of the CG minimisation (seen most clearly in Fig. \ref{fig:fft_asm_ct} between $n=1$ and $n=2$). This is due to a mismatch between the simulation and the experiment, leading to a discrepancy between $\varepsilon_{\text{P}}$ and $\varepsilon_{\text{M}}$ (see Fig. \ref{fig:convergence}). The lowest value of $\varepsilon_{\text{M}}$ is found after less than 100 CG iterations (see hollow circles in the inset of Fig. \ref{fig:fft_asm_ct}).  We did not find a significant improvement by using the ASM instead of the FFT after camera feedback. In optical setups involving a high-NA microscope objective where the paraxial approximation does not hold, we expect the ASM to perform better than in our test setup. 

\subsection{Comparing light potentials}
\label{sec:comp_potentials}

\captionsetup[figure]{aboveskip=3pt}
\begin{figure}[!b]
\centering{\phantomsubcaption\label{fig:ring}\phantomsubcaption\label{fig:harm_conf}\phantomsubcaption\label{fig:spot_array}\phantomsubcaption\label{fig:or-gate}}
\includegraphics[width=14cm]{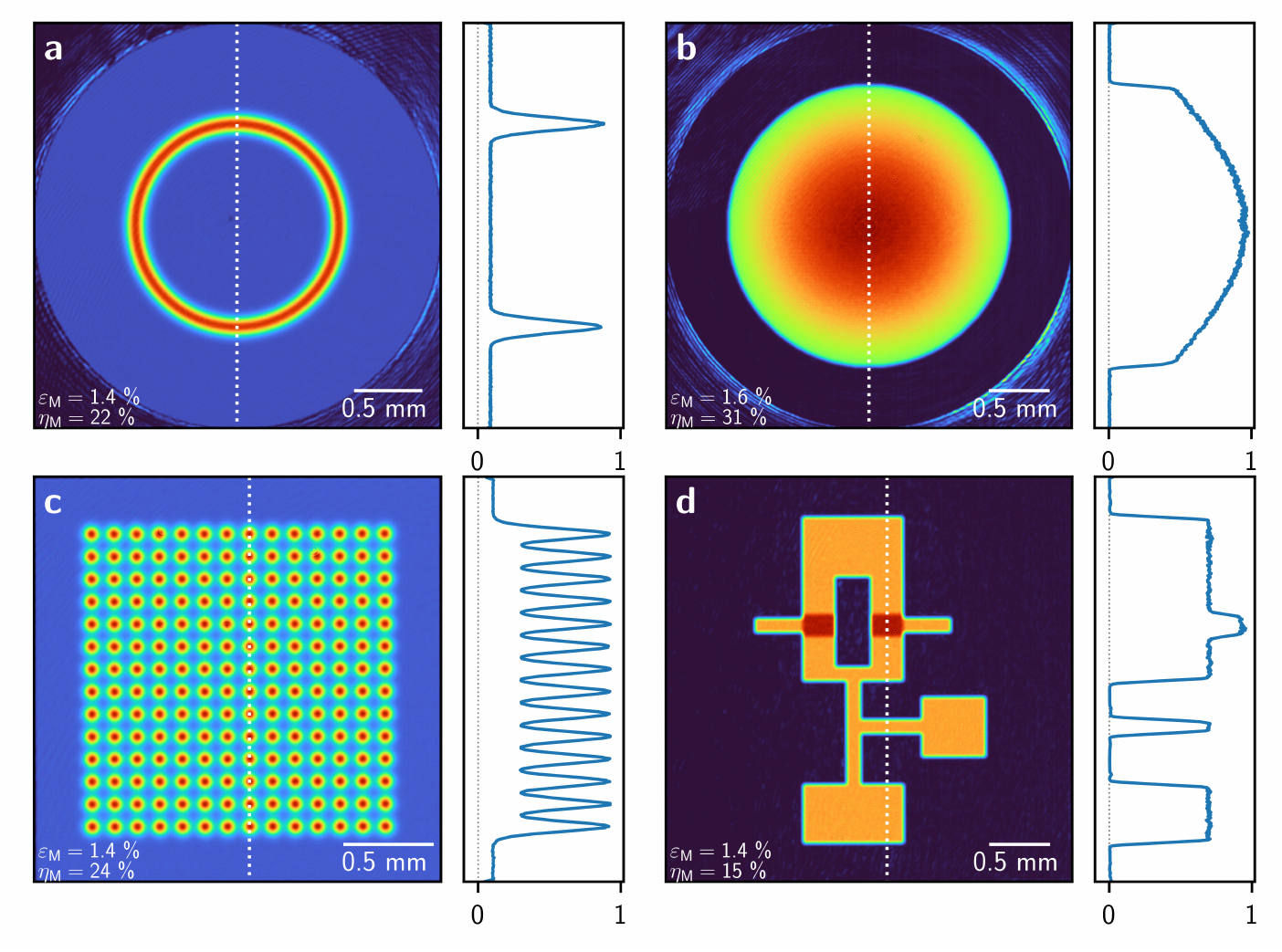}
\caption{Camera images and their normalised profiles (along the white dashed lines) after 15 feedback iterations using the FFT with the crosstalk model. (\subref{fig:ring}) Ring with a Gaussian profile on a non-zero background. (\subref{fig:harm_conf}) Gaussian potential with offset. (\subref{fig:spot_array}) Gaussian spot array on a non-zero background. (\subref{fig:or-gate}) An `atomtronic' logical OR gate \cite{DeMarco2008}.}
\label{fig:light_potentials}
\end{figure}

To characterise our method, we produced various light potentials for cold atom experiments.
We created a ring with a Gaussian profile relevant for atomtronic experiments \cite{Wright2013} (Fig. \ref{fig:ring}), a Gaussian potential with an offset as used to cancel the harmonic confinement in optical lattices \cite{Mazurenko2017}  (Fig. \ref{fig:harm_conf}) and a Gaussian spot array with a non-zero background for tweezer arrays \cite{Ebadi2021} (Fig. \ref{fig:spot_array}).
We also generated a potential resembling an `atomtronic' OR gate as used by previous studies \cite{DeMarco2008, Gaunt2012, VanBijnen2013} (Fig. \ref{fig:or-gate}).
For the Gaussian potential and the spot array, we achieved the best experimental results by using an initial phase guess according to equation \ref{eq:guess} (see \nameref{sec:methods}).
For the ring-shaped potential (Fig. \ref{fig:ring}) and the OR gate (Fig. \ref{fig:or-gate}), an initial phase guess resulting in vortex-free potentials could not be found in the same way.
For these patterns, remaining optical vortices were removed \cite{Sent2018} (see \nameref{sec:methods}).

We find that the vortex removal process introduces high-frequency components in the displayed SLM phase $\varphi\!\left(x, y\right)$ which increases the effect of pixel crosstalk and deteriorates the experimental light potential.
Using our technique, vortex-free simulated light potentials can be generated even from an entirely random initial guess.
However, starting with such a random initial phase guess results in less accurate experimental light potentials.
For all patterns, we used 15 feedback iterations with 100 CG iterations each, accounting for pixel crosstalk during the optimisation. The experimental RMS error of all four patterns varies between $\varepsilon_{\text{M}}=1.4\%-1.6\%$, with measured efficiencies between $\eta_{\text{M}}=15\%-31\%$ (see Table \ref{tab:lit}). The remaining imperfections are most visible in the profiles of light potentials with flat regions. The peak signal-to-noise ratios (PSNR) \cite{Clark2016} measured in the transformed signal region, $S_U$, of the light potentials in Fig. \ref{fig:ring}-\subref{fig:or-gate} are $\unit[45.7]{dB}$, $\unit[40.7]{dB}$, $\unit[43.8]{dB}$, and  $\unit[39.9]{dB}$, respectively.

\section{Discussion}
Compared to previous studies (Table \ref{tab:lit}), we can generate experimental light potentials of low RMS error and higher efficiencies.
Using the CG method, a small line-shaped potential ( 
 $\unit[105]{\mu m}$  length) of $0.7\%$ RMS error has been generated \cite{Ebadi2021} by optimising the intensity and phase in the image plane as well as the efficiency ($\varepsilon_{\text{P}}=38\%$).
Using such a phase constraint, it becomes increasingly difficult to generate light potentials which are accurate and efficient for larger patterns.
A larger line-shaped potential ($\sim  \unit[400]{\mu m}$ length) has been generated in a different study \cite{Bowman2017, Bowman2018} by constraining the phase, however, with a much lower efficiency of $8.3\%$.
If the phase of the target light potential is constrained, more accurate light potentials are typically less efficient and vice versa \cite{Bowman2017, Ebadi2021}.
By removing the phase constraint, accurate and efficient light potentials have been generated computationally using the CG method \cite{Harte2014}, however, the unrestrained phase makes it difficult to realise these experimentally \cite{Bowman2018}.
In this work, we minimised experimental errors by characterising our optical system and by using camera feedback.
This allows us to generate accurate and efficient light potentials experimentally, without constraining the phase in the image plane.
Previous studies have characterised their optical system and used camera feedback without constraining the phase \cite{Gaunt2012, Bruce2015}, however, using an IFTA (MRAF or OMRAF) instead of the CG algorithm, resulting in less accurate and less efficient experimental light potentials than presented here.
In our work, accounting for pixel crosstalk further reduced the RMS error, especially for large light potentials, while lowering the efficiency by $\sim 20\%$ (see bottom of Table \ref{tab:lit}). 

We did not see an improvement in the RMS error when using the ASM instead of the FFT, however, other experimental uncertainties such as a displacement of the Fourier lens in the $xy$-plane or a tilt of the Fourier lens could be modelled with the ASM to improve the accuracy of the light potentials before any camera feedback. Cold atom experiments require microscopic potentials to be projected using a high-NA objective, which will be the subject of further work. The FFT might not be sufficient to model this high-NA objective due to the large diffraction angles and the ASM could lead to more accurate potentials in this scenario, even without restricting the phase.

\begin{table}[!b]
\centering
    \begin{tabular}{||p{1.85cm}p{3.9cm}cc|cc|cc||}
        \hline
        \multicolumn{4}{||c|}{} & \multicolumn{2}{c|}{Simulation} & \multicolumn{2}{c||}{Experiment} \\
        Publication & Pattern & Method & Propagation & $\varepsilon_{\text{P}}$ [\%] & $\eta_{\text{P}}$ [\%] & $\varepsilon_{\text{M}}$ [\%] & $\eta_{\text{M}}$ [\%]\\
        \hline\hline
        Ebadi et al. \cite{Ebadi2021}       & Gaussian line                                     & CG        & FFT   & -     & 38    & 0.7   & -\\
        Bowman \cite{Bowman2018}            & Gaussian line                                     & CG        & FFT   & 0.5   & 8.3   & -     & 3.5\\
        Bruce et al. \cite{Bruce2011}       & Gaussian ring                                     & MRAF      & FFT   & 0.6   & -     & 3.9   & -\\
        This work                           & Gaussian ring (Fig. \ref{fig:ring})               & CG (CT)   & FFT   & 0.56  & 34    & 1.4   & 22\\
        Gaunt et al. \cite{Gaunt2012}       & OR gate                                           & OMRAF     & ASM   & 1     & 24    & 7     & -\\
        Van Bijnen \cite{VanBijnen2013}     & OR gate                                           & MRAF      & FFT   & -     & -     & 6     & -\\
        This work                           & OR gate (Fig. \ref{fig:or-gate})                  & CG (CT)   & FFT   & 0.81  & 24    & 1.4   & 15\\
        Harte et al. \cite{Harte2014}       & Power-law potential                               & CG        & FFT   & 0.07  & 64    & -     & -\\
        This work                           & Gaussian potential (Fig. \ref{fig:harm_conf})     & CG (CT)   & FFT   & 0.80  & 55    & 1.6   & 31\\
        Gaunt et al. \cite{Gaunt2012}       & Top-hat                                           & OMRAF     & ASM   & -     & -     & 6     & -\\
        Van Bijnen \cite{VanBijnen2013}     & Top-hat                                           & MRAF      & FFT   & -     & -     & 1.7   & -\\
        This work                           & Small disc (Fig. \ref{fig:disc_small})            & CG        & FFT   & 0.91  & 87    & 1.1   & 40\\
        This work                           & Spot array (Fig. \ref{fig:spot_array})            & CG (CT)   & FFT   & 0.74  & 41    & 1.4   & 24\\
        This work                           & Large disc (Fig. \ref{fig:fft_asm_ct})                                       & CG        & FFT   & 1.1   & 78    & 2.7   & 34\\
        This work                           & Large disc (Fig. \ref{fig:fft_asm_ct})                                        & CG        & ASM   & 1.0   & 67    & 2.8   & 33\\
        This work                           & Large disc (Fig. \ref{fig:fft_asm_ct})                                        & CG (CT)   & FFT   & 0.92  & 54    & 1.9   & 28\\
        This work                           & Large disc (Fig. \ref{fig:fft_asm_ct})                                        & CG (CT)   & ASM   & 0.87  & 55    & 2.1   & 27\\
        \hline
    \end{tabular}
    \caption{Simulated and experimental errors and efficiencies of previous studies compared to this work. In the last four rows, we compare different methods using the disc-shaped target light potential (convergence shown in Fig. \ref{fig:fft_asm_ct}).}
    \label{tab:lit}
\end{table}

\section{Methods}
\label{sec:methods}
\subsection*{Phase retrieval problem}
\label{sec:phase_retrieval}
The electric field in the SLM plane at $z=0$, $E\!\left(x, y, 0\right)\equiv E_{\text{SLM}}\!\left(x, y\right)$, is calculated using the amplitude of the incident laser beam, $A_{\text{SLM}}\!\left(x, y\right)$, and the phase at the SLM (see Fig.\,\ref{fig:algorithm})
\begin{equation}
E_{\text{SLM}}\!\left(x, y\right) = A_{\text{SLM}}\!\left(x, y\right)\exp{\bigl\{ i\left[ \varphi_{\text{C}}\!\left(x, y\right) + \varphi\!\left(x, y\right)\right]\bigr\}}.
\label{eq:e_slm}
\end{equation}
The phase at the SLM is the sum of the pattern displayed by the SLM, $\varphi\!\left(x, y\right)$, and a constant phase, $\varphi_{\text{C}}\!\left(x, y\right)$, which varies spatially across the SLM but does not change with the displayed phase pattern. This constant phase is caused by distortions of the incoming wavefront and imperfections of the SLM surface. In the image plane at $z=2f$, the electric field, $E\!\left(x, y, 2f\right)\equiv E_{\text{IMG}}\!\left(x, y\right)$, is characterised by the amplitude, $A_{\text{IMG}}\!\left(x, y\right)$, and the phase, $\phi\!\left(x, y\right)$, of the light potential
\begin{equation}
E_{\text{IMG}}\!\left(x, y\right) = A_{\text{IMG}}\!\left(x, y\right)\exp{\left[ i\phi\!\left(x, y\right)\right]}.
\label{eq:e_img}
\end{equation}
Under the paraxial approximation and the far-field approximation, the electric field in the image plane is related to the electric field in the SLM plane via the Fourier transform \cite{Goodman2017}, $\mathcal{F}$, 

\begin{equation}
E_{\text{IMG}}\!\left(\kappa_x, \kappa_y\right)= \frac{1}{i\lambda f}\iint^{\infty}_{-\infty}E_{\text{SLM}}\!\left(x', y'\right)\exp{\left[-2\pi i\left(\kappa_xx' + \kappa_yy'\right)\right]}\,dx'\,dy' \equiv\mathcal{F}\left\{ E_{\text{SLM}}\left(x, y\right)\right\},
\label{eq:fourier}
\end{equation}
with spatial frequencies in the image plane, $\kappa_x=x/\lambda f$ and $\kappa_y=y/\lambda f$.

\subsection*{Implementation of conjugate gradient minimisation and camera feedback}
\label{sec:methods_cg}
We use a nonlinear CG solver \cite{polak1969}, implemented on a GPU using PyTorch which has automatic differentiation capabilities.
This allows us to compute the gradient of the cost function, $\partial C / \partial \varphi$, without the need for an analytic expression.
Using $s=10^{12}$ (see equation \ref{eq:cost}), the minimisation typically reaches $\varepsilon_{\text{P}}=1\%$ within 100 iterations, depending on the shape of the desired potential and provided that an initial phase guess which does not lead to optical vortices was used. As the SLM phase pattern is optimised by simulating the diffraction of light, the target intensity pattern, $\tilde{T}_{kl}$, is convolved with the point spread function of our optical system to remove sub-diffraction limited features which hinder convergence. If desired, a term could be added to the cost function to optimise for a higher power inside the signal region \cite{Ebadi2021}. Currently, we do not require control over the phase, $\phi_{kl}$, in the image plane, however, it is possible to simultaneously control the intensity and the phase in the image plane at the expense of efficiency \cite{Bowman2018, Ebadi2021}.

The camera feedback algorithm \cite{Bruce2015} further improves the CG hologram calculation. 
Initially, an SLM phase pattern, $\varphi_{ij}^{\left(0\right)}$, is calculated for a given target light potential, $\hat{T}_{kl}^{\left(0\right)}$, by running the CG minimisation for $m_{\text{max}}$ iterations. We then display this pattern on the SLM and take a camera image, $I_{uv}$, of the light potential. We map the initial target light potential from the coordinate system of the computational image plane, $\hat{T}^{(0)}_{kl}$, to the coordinate system of the camera image, $\hat{T}^{(0)}_{uv}$, using an affine transformation.
To calculate the affine transformation, we generate a checkerboard-shaped light potential using the CG algorithm and detect the corner points of the checkerboard in the camera image \cite{Geiger2012, Peng2020}.
Then, the camera image, $I_{uv}$, and the transformed initial target light potential, $T^{(0)}_{uv}$, are normalised \cite{Bruce2015} and subtracted from each other. 
This difference $D_{uv}=\hat{T}^{(0)}_{uv} - \hat{I}_{uv}$ is then transformed back to the coordinates of the computational image plane and added to the previous target light potential $\hat{T}^{(n-1)}_{kl}$, resulting in a new target light potential $\hat{T}^{(n)}_{kl}=\hat{T}^{(n-1)}_{kl} + D_{kl}$ for the next feedback iteration.
We then re-run the CG minimisation using the updated target light potential and the previous optimised phase pattern, $\varphi_{ij}^{\left(n-1\right)}$, as an initial guess.
Before the new target potential is calculated, the difference $D_{uv}$ is blurred with a Gaussian kernel to ensure that there are no features in the new target that are smaller than the diffraction limit (e.g. camera noise) as the CG minimisation cannot produce light potentials containing sub-diffraction-limited features.

\phantomsection
\subsection{Initial phase guess}
\label{sec:guess}

\begin{figure}[!tb] \centering{\phantomsubcaption\label{fig:int_vtx}\phantomsubcaption\label{fig:phi_vtx}\phantomsubcaption\label{fig:phi_avtx}\phantomsubcaption\label{fig:phi_corr}\phantomsubcaption\label{fig:n_vtx}\phantomsubcaption\label{fig:rmse_vtx}}
    \includegraphics[width=14cm]{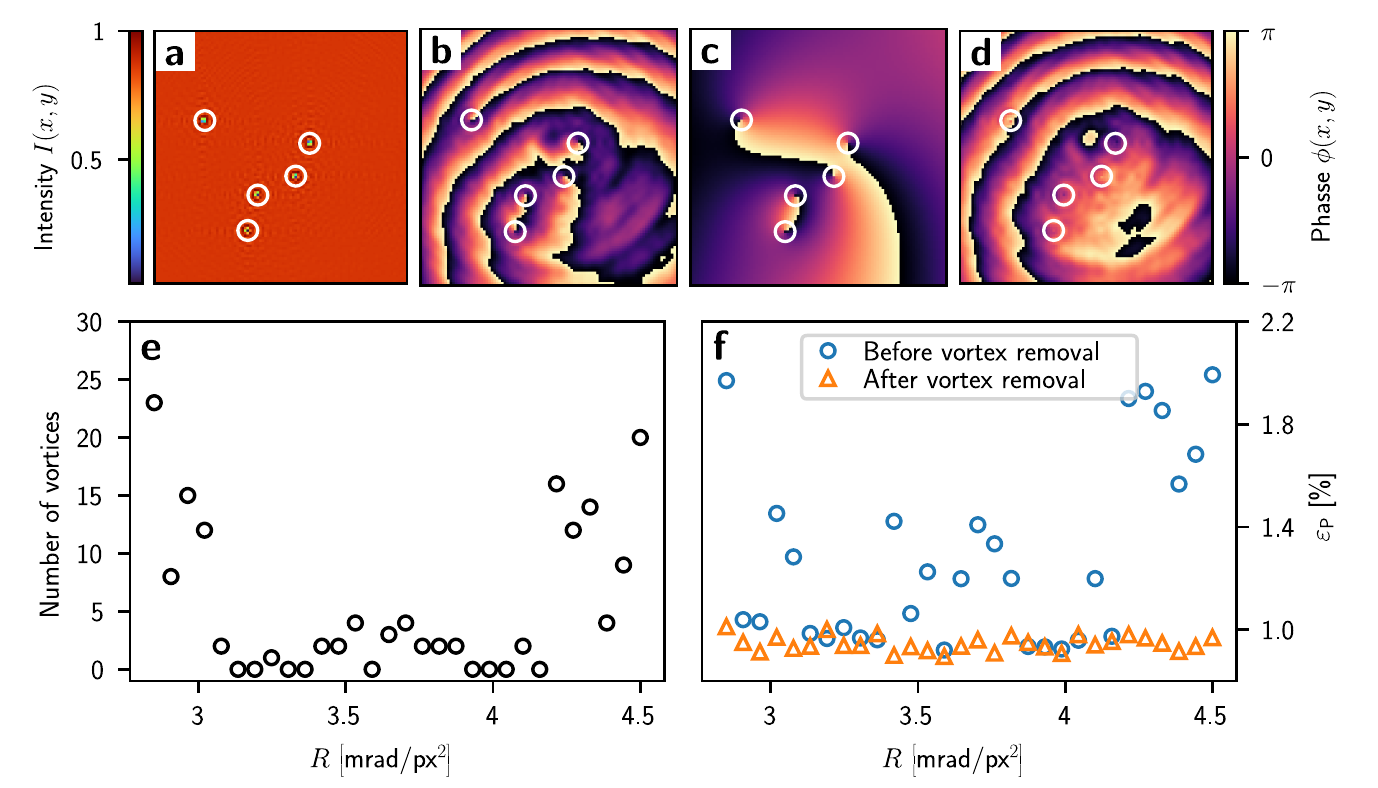}
    \caption{Detection and removal of optical vortices in the disc-shaped light potential. (\subref{fig:int_vtx}) Intensity of  the light potential, showing the central $100 \times 100$ pixels;
    (\subref{fig:phi_vtx}) phase,  $\phi$, of the same potential, (\subref{fig:phi_avtx}) phase, $\phi_{\text{v}}$ of the vortices only, (\subref{fig:phi_corr})  phase, $\phi - \phi_{\text{v}}$, of the corrected field with vortices removed.
    (\subref{fig:n_vtx}) Number of vortices detected in the light potential after 100 CG iterations and 10 feedback iterations, using different values for the quadratic curvature, $R$, in the initial phase guess.  (\subref{fig:rmse_vtx}) Predicted RMS error, $\varepsilon_{\text{P}}$, before vortex removal (blue circles) and after (orange triangles).}
\label{fig:guess}
\end{figure}

We use a combination of a linear phase and a quadratic phase as an initial phase guess, $\varphi_{\text{G}}$, which is common practice in IFTAs and gradient-based phase retrieval algorithms \cite{DeMarco2008, Harte2014},
\begin{equation}
\varphi_{\text{G}}\left( x, y\right) = m_{x}x + m_{y}y + 4R\left[ \gamma x^2 + \left(1-\gamma\right) y^2\right],
\label{eq:guess}
\end{equation}
The linear terms $m_{x}x$ and $m_{y}y$ diffract the light away from the optical axis and are typically determined by the shape of the target light potential, $T_{kl}$.
The quadratic term with curvature, $R$, and aspect ratio, $\gamma$, are used to control the size of the illuminated area.
Smaller values of $R$ produce more efficient light potentials as more light is focused into the signal region $S$.
The initial phase guess must be chosen such that optical vortices cannot form in the signal region $S$ of the image plane \cite{DeMarco2008, Harte2014}.
An optical vortex is a phase winding around a singularity at which the phase is not defined \cite{Sent2018}.
The field amplitude at this point is zero, causing `holes' in the light potential (Fig. \ref{fig:int_vtx}).
The CG minimisation cannot remove these vortices because a global phase shift is required to annihilate them \cite{Schimmel2005}.
By varying $R$, an initial guess that prevents the formation of optical vortices can be found for `simple' target potentials.
We choose a uniform disc on a dark background as a target potential and detect the number of vortices in the resulting light potential for each value of $R$ (Fig. \ref{fig:n_vtx}).
The vortices in the light potential cause a higher predicted RMS error, $\varepsilon_{\text{P}}$ (black circles in Fig. \ref{fig:n_vtx} and blue circles in Fig. \ref{fig:rmse_vtx}).
Certain values of $R$ do not result in optical vortices, and the lowest RMS error was found for $R=\unit[3.6]{mrad/\text{px}^{2}} $.

This procedure works well for simple patterns such as a disc-shaped flattop, however, for more intricate light potentials, it becomes difficult to find a suitable initial guess by scanning the value of $R$. Further, we found that using the measured intensity profile of the incident laser beam, $\left|A_{\text{SLM}}\left(x, y\right)\right|^2$, instead of a perfect Gaussian can introduce vortices even for simple patterns. To improve our scheme, we detect optical vortices in the light potential and remove them \cite{Sent2018, Schimmel2005}. Initially, the usual CG minimisation is performed until stagnation is reached. We then detect the position of the vortices by identifying the zero crossings of the real and imaginary part of the electric field in the image plane, $E_{\text{IMG}}\!\left(x, y\right)$. To find the charge of the vortices, a line integral around the $3 \times 3$ neighbours of these points is evaluated. The sign of the line integral indicates if the vortex is of positive or negative charge. The phase around these vortices, $\phi_{\text{V}}\!\left(x, y\right)$, is calculated using the relation \cite{Schimmel2005}
\begin{equation}
\phi_{\text{V}}\left(x, y\right) = \sum_{n=1}^{N}q_{n}\Arg\left[\left(x-x_{n}\right)+i\left(y-y_{n}\right)\right],
\label{eq:annihil}
\end{equation}
where $N$ is the total number of vortices, $q_{n}$ the charge of the vortex and $x_{n}$ and $y_{n}$ its position. The phase, $\phi_{\text{V}}\!\left(x, y\right)$, (Fig. \ref{fig:phi_avtx}) is then subtracted from the phase of the light potential, $\phi\!\left(x, y\right)$, (Fig. \ref{fig:phi_vtx}) which annihilates the vortices (Fig. \ref{fig:phi_corr}). The electric field consisting of the corrected phase, $\phi\!\left(x, y\right) - \phi_{\text{V}}\!\left(x, y\right)$, and the amplitude of the light potential, $A_{\text{IMG}}\!\left(x, y\right)$, is propagated back to the SLM plane using the inverse Fourier transform. The phase of the resulting electric field is used as a new initial phase guess, $\varphi_{\text{G}}\!\left(x, y\right)$,
\begin{equation}
\varphi_{\text{G}}\!\left(x, y\right) = \Arg\biggl[\mathcal{F}^{-1}\Bigl\{ A_{\text{IMG}}\!\left(x, y\right)\exp\bigl[ i\left( \phi\!\left(x, y\right) - \phi_{\text{V}}\!\left(x, y\right)\right)\bigr]\Bigr\}\biggr].
\label{eq:vtx_guess}
\end{equation}
By re-running the CG minimisation using $\varphi_{\text{G}}\!\left(x, y\right)$, a vortex-free light potential can be produced, provided that all vortices in the light potential were detected. In case there are remaining vortices in the light potential, this process can be repeated until all vortices are detected and annihilated.

\subsection{Efficiency measurement}
To obtain the power in the signal region, $P_S$, we measure the optical power that corresponds to a certain pixel value and exposure time of the camera image. We display a circular mask on the SLM containing a linear phase gradient and place an iris in the image plane to block the zeroth-order light. Only the power of the first-order spot caused by the SLM phase pattern is measured using a power meter. We then take a camera image of this spot with a certain exposure time and relate the pixel sum of the camera image to the measured power. Using this calibration, the optical power, $P_S$, is calculated from the pixel sum of the camera image inside the transformed signal region, $\sum_{u,v\in S_U}I_{uv}$, and the exposure time. The predicted efficiency, $\eta_{\text{P}}$, is always higher than the measured efficiency, $\eta_{\text{M}}$, as it does not take the diffraction efficiency of the SLM into account. When displaying a flat phase on the SLM, the measured power of the zeroth-order spot is $69\%$ of the incident power, $P_{\text{in}}$.


\bibliography{sample}

\begin{thebibliography}{42}%
\makeatletter
\providecommand \@ifxundefined [1]{%
 \@ifx{#1\undefined}
}%
\providecommand \@ifnum [1]{%
 \ifnum #1\expandafter \@firstoftwo
 \else \expandafter \@secondoftwo
 \fi
}%
\providecommand \@ifx [1]{%
 \ifx #1\expandafter \@firstoftwo
 \else \expandafter \@secondoftwo
 \fi
}%
\providecommand \natexlab [1]{#1}%
\providecommand \enquote  [1]{``#1''}%
\providecommand \bibnamefont  [1]{#1}%
\providecommand \bibfnamefont [1]{#1}%
\providecommand \citenamefont [1]{#1}%
\providecommand \href@noop [0]{\@secondoftwo}%
\providecommand \href [0]{\begingroup \@sanitize@url \@href}%
\providecommand \@href[1]{\@@startlink{#1}\@@href}%
\providecommand \@@href[1]{\endgroup#1\@@endlink}%
\providecommand \@sanitize@url [0]{\catcode `\\12\catcode `\$12\catcode
  `\&12\catcode `\#12\catcode `\^12\catcode `\_12\catcode `\%12\relax}%
\providecommand \@@startlink[1]{}%
\providecommand \@@endlink[0]{}%
\providecommand \url  [0]{\begingroup\@sanitize@url \@url }%
\providecommand \@url [1]{\endgroup\@href {#1}{\urlprefix }}%
\providecommand \urlprefix  [0]{URL }%
\providecommand \Eprint [0]{\href }%
\providecommand \doibase [0]{https://doi.org/}%
\providecommand \selectlanguage [0]{\@gobble}%
\providecommand \bibinfo  [0]{\@secondoftwo}%
\providecommand \bibfield  [0]{\@secondoftwo}%
\providecommand \translation [1]{[#1]}%
\providecommand \BibitemOpen [0]{}%
\providecommand \bibitemStop [0]{}%
\providecommand \bibitemNoStop [0]{.\EOS\space}%
\providecommand \EOS [0]{\spacefactor3000\relax}%
\providecommand \BibitemShut  [1]{\csname bibitem#1\endcsname}%
\let\auto@bib@innerbib\@empty
\bibitem [{\citenamefont {Kwon}\ \emph {et~al.}(2020)\citenamefont {Kwon},
  \citenamefont {Pace}, \citenamefont {Panza}, \citenamefont {Inguscio},
  \citenamefont {Zwerger}, \citenamefont {Zaccanti}, \citenamefont {Scazza},\
  and\ \citenamefont {Roati}}]{Kwon2020}%
  \BibitemOpen
  \bibfield  {author} {\bibinfo {author} {\bibfnamefont {W.~J.}\ \bibnamefont
  {Kwon}}, \bibinfo {author} {\bibfnamefont {G.~D.}\ \bibnamefont {Pace}},
  \bibinfo {author} {\bibfnamefont {R.}~\bibnamefont {Panza}}, \bibinfo
  {author} {\bibfnamefont {M.}~\bibnamefont {Inguscio}}, \bibinfo {author}
  {\bibfnamefont {W.}~\bibnamefont {Zwerger}}, \bibinfo {author} {\bibfnamefont
  {M.}~\bibnamefont {Zaccanti}}, \bibinfo {author} {\bibfnamefont
  {F.}~\bibnamefont {Scazza}},\ and\ \bibinfo {author} {\bibfnamefont
  {G.}~\bibnamefont {Roati}},\ }\bibfield  {title} {\bibinfo {title} {{Strongly
  correlated superfluid order parameters from dc Josephson supercurrents}},\
  }\href {https://doi.org/https://doi.org/10.1126/science.aaz2463} {\bibfield
  {journal} {\bibinfo  {journal} {Science}\ }\textbf {\bibinfo {volume}
  {369}},\ \bibinfo {pages} {84} (\bibinfo {year} {2020})}\BibitemShut
  {NoStop}%
\bibitem [{\citenamefont {Luick}\ \emph {et~al.}(2020)\citenamefont {Luick},
  \citenamefont {Sobirey}, \citenamefont {Bohlen}, \citenamefont {Singh},
  \citenamefont {Mathey}, \citenamefont {Lompe},\ and\ \citenamefont
  {Moritz}}]{Luick2020}%
  \BibitemOpen
  \bibfield  {author} {\bibinfo {author} {\bibfnamefont {N.}~\bibnamefont
  {Luick}}, \bibinfo {author} {\bibfnamefont {L.}~\bibnamefont {Sobirey}},
  \bibinfo {author} {\bibfnamefont {M.}~\bibnamefont {Bohlen}}, \bibinfo
  {author} {\bibfnamefont {V.~P.}\ \bibnamefont {Singh}}, \bibinfo {author}
  {\bibfnamefont {L.}~\bibnamefont {Mathey}}, \bibinfo {author} {\bibfnamefont
  {T.}~\bibnamefont {Lompe}},\ and\ \bibinfo {author} {\bibfnamefont
  {H.}~\bibnamefont {Moritz}},\ }\bibfield  {title} {\bibinfo {title} {{An
  ideal Josephson junction in an ultracold two-dimensional Fermi gas}},\ }\href
  {https://doi.org/https://doi.org/10.1126/science.aaz2342} {\bibfield
  {journal} {\bibinfo  {journal} {Science}\ }\textbf {\bibinfo {volume}
  {369}},\ \bibinfo {pages} {89} (\bibinfo {year} {2020})}\BibitemShut
  {NoStop}%
\bibitem [{\citenamefont {Guo}\ \emph {et~al.}(2021)\citenamefont {Guo},
  \citenamefont {Kroeze}, \citenamefont {Marsh}, \citenamefont
  {Gopalakrishnan}, \citenamefont {Keeling},\ and\ \citenamefont
  {Lev}}]{Guo2021}%
  \BibitemOpen
  \bibfield  {author} {\bibinfo {author} {\bibfnamefont {Y.}~\bibnamefont
  {Guo}}, \bibinfo {author} {\bibfnamefont {R.~M.}\ \bibnamefont {Kroeze}},
  \bibinfo {author} {\bibfnamefont {B.~P.}\ \bibnamefont {Marsh}}, \bibinfo
  {author} {\bibfnamefont {S.}~\bibnamefont {Gopalakrishnan}}, \bibinfo
  {author} {\bibfnamefont {J.}~\bibnamefont {Keeling}},\ and\ \bibinfo {author}
  {\bibfnamefont {B.~L.}\ \bibnamefont {Lev}},\ }\bibfield  {title} {\bibinfo
  {title} {An optical lattice with sound},\ }\href
  {https://doi.org/https://doi.org/10.1038/s41586-021-03945-x} {\bibfield
  {journal} {\bibinfo  {journal} {Nature}\ }\textbf {\bibinfo {volume} {599}},\
  \bibinfo {pages} {211} (\bibinfo {year} {2021})}\BibitemShut {NoStop}%
\bibitem [{\citenamefont {Sompet}\ \emph {et~al.}(2022)\citenamefont {Sompet},
  \citenamefont {Hirthe}, \citenamefont {Bourgund}, \citenamefont {Chalopin},
  \citenamefont {Bibo}, \citenamefont {Koepsell}, \citenamefont {Bojović},
  \citenamefont {Verresen}, \citenamefont {Pollmann}, \citenamefont {Salomon},
  \citenamefont {Gross}, \citenamefont {Hilker},\ and\ \citenamefont
  {Bloch}}]{Sompet2022}%
  \BibitemOpen
  \bibfield  {author} {\bibinfo {author} {\bibfnamefont {P.}~\bibnamefont
  {Sompet}}, \bibinfo {author} {\bibfnamefont {S.}~\bibnamefont {Hirthe}},
  \bibinfo {author} {\bibfnamefont {D.}~\bibnamefont {Bourgund}}, \bibinfo
  {author} {\bibfnamefont {T.}~\bibnamefont {Chalopin}}, \bibinfo {author}
  {\bibfnamefont {J.}~\bibnamefont {Bibo}}, \bibinfo {author} {\bibfnamefont
  {J.}~\bibnamefont {Koepsell}}, \bibinfo {author} {\bibfnamefont
  {P.}~\bibnamefont {Bojović}}, \bibinfo {author} {\bibfnamefont
  {R.}~\bibnamefont {Verresen}}, \bibinfo {author} {\bibfnamefont
  {F.}~\bibnamefont {Pollmann}}, \bibinfo {author} {\bibfnamefont
  {G.}~\bibnamefont {Salomon}}, \bibinfo {author} {\bibfnamefont
  {C.}~\bibnamefont {Gross}}, \bibinfo {author} {\bibfnamefont {T.~A.}\
  \bibnamefont {Hilker}},\ and\ \bibinfo {author} {\bibfnamefont
  {I.}~\bibnamefont {Bloch}},\ }\bibfield  {title} {\bibinfo {title}
  {{Realizing the symmetry-protected Haldane phase in Fermi–Hubbard
  ladders}},\ }\href
  {https://doi.org/https://doi.org/10.1038/s41586-022-04688-z} {\bibfield
  {journal} {\bibinfo  {journal} {Nature}\ }\textbf {\bibinfo {volume} {606}},\
  \bibinfo {pages} {484} (\bibinfo {year} {2022})}\BibitemShut {NoStop}%
\bibitem [{\citenamefont {Mazurenko}\ \emph {et~al.}(2017)\citenamefont
  {Mazurenko}, \citenamefont {Chiu}, \citenamefont {Ji}, \citenamefont
  {Parsons}, \citenamefont {Kanász-Nagy}, \citenamefont {Schmidt},
  \citenamefont {Grusdt}, \citenamefont {Demler}, \citenamefont {Greif},\ and\
  \citenamefont {Greiner}}]{Mazurenko2017}%
  \BibitemOpen
  \bibfield  {author} {\bibinfo {author} {\bibfnamefont {A.}~\bibnamefont
  {Mazurenko}}, \bibinfo {author} {\bibfnamefont {C.~S.}\ \bibnamefont {Chiu}},
  \bibinfo {author} {\bibfnamefont {G.}~\bibnamefont {Ji}}, \bibinfo {author}
  {\bibfnamefont {M.~F.}\ \bibnamefont {Parsons}}, \bibinfo {author}
  {\bibfnamefont {M.}~\bibnamefont {Kanász-Nagy}}, \bibinfo {author}
  {\bibfnamefont {R.}~\bibnamefont {Schmidt}}, \bibinfo {author} {\bibfnamefont
  {F.}~\bibnamefont {Grusdt}}, \bibinfo {author} {\bibfnamefont
  {E.}~\bibnamefont {Demler}}, \bibinfo {author} {\bibfnamefont
  {D.}~\bibnamefont {Greif}},\ and\ \bibinfo {author} {\bibfnamefont
  {M.}~\bibnamefont {Greiner}},\ }\bibfield  {title} {\bibinfo {title} {{A
  cold-atom Fermi–Hubbard antiferromagnet}},\ }\href
  {https://doi.org/https://doi.org/10.1038/nature22362} {\bibfield  {journal}
  {\bibinfo  {journal} {Nature}\ }\textbf {\bibinfo {volume} {545}},\ \bibinfo
  {pages} {462} (\bibinfo {year} {2017})}\BibitemShut {NoStop}%
\bibitem [{\citenamefont {Navon}\ \emph {et~al.}(2016)\citenamefont {Navon},
  \citenamefont {Gaunt}, \citenamefont {Smith},\ and\ \citenamefont
  {Hadzibabic}}]{Navon2016}%
  \BibitemOpen
  \bibfield  {author} {\bibinfo {author} {\bibfnamefont {N.}~\bibnamefont
  {Navon}}, \bibinfo {author} {\bibfnamefont {A.~L.}\ \bibnamefont {Gaunt}},
  \bibinfo {author} {\bibfnamefont {R.~P.}\ \bibnamefont {Smith}},\ and\
  \bibinfo {author} {\bibfnamefont {Z.}~\bibnamefont {Hadzibabic}},\ }\bibfield
   {title} {\bibinfo {title} {Emergence of a turbulent cascade in a quantum
  gas},\ }\href {https://doi.org/https://doi.org/10.1038/nature20114}
  {\bibfield  {journal} {\bibinfo  {journal} {Nature}\ }\textbf {\bibinfo
  {volume} {539}},\ \bibinfo {pages} {72} (\bibinfo {year} {2016})}\BibitemShut
  {NoStop}%
\bibitem [{\citenamefont {Ebadi}\ \emph {et~al.}(2021)\citenamefont {Ebadi},
  \citenamefont {Wang}, \citenamefont {Levine}, \citenamefont {Keesling},
  \citenamefont {Semeghini}, \citenamefont {Omran}, \citenamefont {Bluvstein},
  \citenamefont {Samajdar}, \citenamefont {Pichler}, \citenamefont {Ho},
  \citenamefont {Choi}, \citenamefont {Sachdev}, \citenamefont {Greiner},
  \citenamefont {Vuletić},\ and\ \citenamefont {Lukin}}]{Ebadi2021}%
  \BibitemOpen
  \bibfield  {author} {\bibinfo {author} {\bibfnamefont {S.}~\bibnamefont
  {Ebadi}}, \bibinfo {author} {\bibfnamefont {T.~T.}\ \bibnamefont {Wang}},
  \bibinfo {author} {\bibfnamefont {H.}~\bibnamefont {Levine}}, \bibinfo
  {author} {\bibfnamefont {A.}~\bibnamefont {Keesling}}, \bibinfo {author}
  {\bibfnamefont {G.}~\bibnamefont {Semeghini}}, \bibinfo {author}
  {\bibfnamefont {A.}~\bibnamefont {Omran}}, \bibinfo {author} {\bibfnamefont
  {D.}~\bibnamefont {Bluvstein}}, \bibinfo {author} {\bibfnamefont
  {R.}~\bibnamefont {Samajdar}}, \bibinfo {author} {\bibfnamefont
  {H.}~\bibnamefont {Pichler}}, \bibinfo {author} {\bibfnamefont {W.~W.}\
  \bibnamefont {Ho}}, \bibinfo {author} {\bibfnamefont {S.}~\bibnamefont
  {Choi}}, \bibinfo {author} {\bibfnamefont {S.}~\bibnamefont {Sachdev}},
  \bibinfo {author} {\bibfnamefont {M.}~\bibnamefont {Greiner}}, \bibinfo
  {author} {\bibfnamefont {V.}~\bibnamefont {Vuletić}},\ and\ \bibinfo
  {author} {\bibfnamefont {M.~D.}\ \bibnamefont {Lukin}},\ }\bibfield  {title}
  {\bibinfo {title} {Quantum phases of matter on a 256-atom programmable
  quantum simulator},\ }\href
  {https://doi.org/https://doi.org/10.1038/s41586-021-03582-4} {\bibfield
  {journal} {\bibinfo  {journal} {Nature}\ }\textbf {\bibinfo {volume} {595}},\
  \bibinfo {pages} {227} (\bibinfo {year} {2021})}\BibitemShut {NoStop}%
\bibitem [{\citenamefont {Barredo}\ \emph {et~al.}(2016)\citenamefont
  {Barredo}, \citenamefont {L\'{e}s\'{e}leuc}, \citenamefont {Lienhard},
  \citenamefont {Lahaye},\ and\ \citenamefont {Browaeys}}]{Barredo2016}%
  \BibitemOpen
  \bibfield  {author} {\bibinfo {author} {\bibfnamefont {D.}~\bibnamefont
  {Barredo}}, \bibinfo {author} {\bibfnamefont {S.~D.}\ \bibnamefont
  {L\'{e}s\'{e}leuc}}, \bibinfo {author} {\bibfnamefont {V.}~\bibnamefont
  {Lienhard}}, \bibinfo {author} {\bibfnamefont {T.}~\bibnamefont {Lahaye}},\
  and\ \bibinfo {author} {\bibfnamefont {A.}~\bibnamefont {Browaeys}},\
  }\bibfield  {title} {\bibinfo {title} {An atom-by-atom assembler of
  defect-free arbitrary two-dimensional atomic arrays},\ }\href
  {https://doi.org/https://doi.org/10.1126/science.aah3778} {\bibfield
  {journal} {\bibinfo  {journal} {Science}\ }\textbf {\bibinfo {volume}
  {354}},\ \bibinfo {pages} {1021} (\bibinfo {year} {2016})}\BibitemShut
  {NoStop}%
\bibitem [{\citenamefont {Amico}\ \emph {et~al.}(2021)\citenamefont {Amico},
  \citenamefont {Boshier}, \citenamefont {Birkl}, \citenamefont {Minguzzi},
  \citenamefont {Miniatura}, \citenamefont {Kwek}, \citenamefont {Aghamalyan},
  \citenamefont {Ahufinger}, \citenamefont {Anderson}, \citenamefont {Andrei},
  \citenamefont {Arnold}, \citenamefont {Baker}, \citenamefont {Bell},
  \citenamefont {Bland}, \citenamefont {Brantut}, \citenamefont {Cassettari},
  \citenamefont {Chetcuti}, \citenamefont {Chevy}, \citenamefont {Citro},
  \citenamefont {Palo}, \citenamefont {Dumke}, \citenamefont {Edwards},
  \citenamefont {Folman}, \citenamefont {Fortagh}, \citenamefont {Gardiner},
  \citenamefont {Garraway}, \citenamefont {Gauthier}, \citenamefont {Günther},
  \citenamefont {Haug}, \citenamefont {Hufnagel}, \citenamefont {Keil},
  \citenamefont {Ireland}, \citenamefont {Lebrat}, \citenamefont {Li},
  \citenamefont {Longchambon}, \citenamefont {Mompart}, \citenamefont {Morsch},
  \citenamefont {Naldesi}, \citenamefont {Neely}, \citenamefont {Olshanii},
  \citenamefont {Orignac}, \citenamefont {Pandey}, \citenamefont
  {Pérez-Obiol}, \citenamefont {Perrin}, \citenamefont {Piroli}, \citenamefont
  {Polo}, \citenamefont {Pritchard}, \citenamefont {Proukakis}, \citenamefont
  {Rylands}, \citenamefont {Rubinsztein-Dunlop}, \citenamefont {Scazza},
  \citenamefont {Stringari}, \citenamefont {Tosto}, \citenamefont
  {Trombettoni}, \citenamefont {Victorin}, \citenamefont {Klitzing},
  \citenamefont {Wilkowski}, \citenamefont {Xhani},\ and\ \citenamefont
  {Yakimenko}}]{Amico2021}%
  \BibitemOpen
  \bibfield  {author} {\bibinfo {author} {\bibfnamefont {L.}~\bibnamefont
  {Amico}}, \bibinfo {author} {\bibfnamefont {M.}~\bibnamefont {Boshier}},
  \bibinfo {author} {\bibfnamefont {G.}~\bibnamefont {Birkl}}, \bibinfo
  {author} {\bibfnamefont {A.}~\bibnamefont {Minguzzi}}, \bibinfo {author}
  {\bibfnamefont {C.}~\bibnamefont {Miniatura}}, \bibinfo {author}
  {\bibfnamefont {L.~C.}\ \bibnamefont {Kwek}}, \bibinfo {author}
  {\bibfnamefont {D.}~\bibnamefont {Aghamalyan}}, \bibinfo {author}
  {\bibfnamefont {V.}~\bibnamefont {Ahufinger}}, \bibinfo {author}
  {\bibfnamefont {D.}~\bibnamefont {Anderson}}, \bibinfo {author}
  {\bibfnamefont {N.}~\bibnamefont {Andrei}}, \bibinfo {author} {\bibfnamefont
  {A.~S.}\ \bibnamefont {Arnold}}, \bibinfo {author} {\bibfnamefont
  {M.}~\bibnamefont {Baker}}, \bibinfo {author} {\bibfnamefont {T.~A.}\
  \bibnamefont {Bell}}, \bibinfo {author} {\bibfnamefont {T.}~\bibnamefont
  {Bland}}, \bibinfo {author} {\bibfnamefont {J.~P.}\ \bibnamefont {Brantut}},
  \bibinfo {author} {\bibfnamefont {D.}~\bibnamefont {Cassettari}}, \bibinfo
  {author} {\bibfnamefont {W.~J.}\ \bibnamefont {Chetcuti}}, \bibinfo {author}
  {\bibfnamefont {F.}~\bibnamefont {Chevy}}, \bibinfo {author} {\bibfnamefont
  {R.}~\bibnamefont {Citro}}, \bibinfo {author} {\bibfnamefont {S.~D.}\
  \bibnamefont {Palo}}, \bibinfo {author} {\bibfnamefont {R.}~\bibnamefont
  {Dumke}}, \bibinfo {author} {\bibfnamefont {M.}~\bibnamefont {Edwards}},
  \bibinfo {author} {\bibfnamefont {R.}~\bibnamefont {Folman}}, \bibinfo
  {author} {\bibfnamefont {J.}~\bibnamefont {Fortagh}}, \bibinfo {author}
  {\bibfnamefont {S.~A.}\ \bibnamefont {Gardiner}}, \bibinfo {author}
  {\bibfnamefont {B.~M.}\ \bibnamefont {Garraway}}, \bibinfo {author}
  {\bibfnamefont {G.}~\bibnamefont {Gauthier}}, \bibinfo {author}
  {\bibfnamefont {A.}~\bibnamefont {Günther}}, \bibinfo {author}
  {\bibfnamefont {T.}~\bibnamefont {Haug}}, \bibinfo {author} {\bibfnamefont
  {C.}~\bibnamefont {Hufnagel}}, \bibinfo {author} {\bibfnamefont
  {M.}~\bibnamefont {Keil}}, \bibinfo {author} {\bibfnamefont {P.}~\bibnamefont
  {Ireland}}, \bibinfo {author} {\bibfnamefont {M.}~\bibnamefont {Lebrat}},
  \bibinfo {author} {\bibfnamefont {W.}~\bibnamefont {Li}}, \bibinfo {author}
  {\bibfnamefont {L.}~\bibnamefont {Longchambon}}, \bibinfo {author}
  {\bibfnamefont {J.}~\bibnamefont {Mompart}}, \bibinfo {author} {\bibfnamefont
  {O.}~\bibnamefont {Morsch}}, \bibinfo {author} {\bibfnamefont
  {P.}~\bibnamefont {Naldesi}}, \bibinfo {author} {\bibfnamefont {T.~W.}\
  \bibnamefont {Neely}}, \bibinfo {author} {\bibfnamefont {M.}~\bibnamefont
  {Olshanii}}, \bibinfo {author} {\bibfnamefont {E.}~\bibnamefont {Orignac}},
  \bibinfo {author} {\bibfnamefont {S.}~\bibnamefont {Pandey}}, \bibinfo
  {author} {\bibfnamefont {A.}~\bibnamefont {Pérez-Obiol}}, \bibinfo {author}
  {\bibfnamefont {H.}~\bibnamefont {Perrin}}, \bibinfo {author} {\bibfnamefont
  {L.}~\bibnamefont {Piroli}}, \bibinfo {author} {\bibfnamefont
  {J.}~\bibnamefont {Polo}}, \bibinfo {author} {\bibfnamefont {A.~L.}\
  \bibnamefont {Pritchard}}, \bibinfo {author} {\bibfnamefont {N.~P.}\
  \bibnamefont {Proukakis}}, \bibinfo {author} {\bibfnamefont {C.}~\bibnamefont
  {Rylands}}, \bibinfo {author} {\bibfnamefont {H.}~\bibnamefont
  {Rubinsztein-Dunlop}}, \bibinfo {author} {\bibfnamefont {F.}~\bibnamefont
  {Scazza}}, \bibinfo {author} {\bibfnamefont {S.}~\bibnamefont {Stringari}},
  \bibinfo {author} {\bibfnamefont {F.}~\bibnamefont {Tosto}}, \bibinfo
  {author} {\bibfnamefont {A.}~\bibnamefont {Trombettoni}}, \bibinfo {author}
  {\bibfnamefont {N.}~\bibnamefont {Victorin}}, \bibinfo {author}
  {\bibfnamefont {W.~V.}\ \bibnamefont {Klitzing}}, \bibinfo {author}
  {\bibfnamefont {D.}~\bibnamefont {Wilkowski}}, \bibinfo {author}
  {\bibfnamefont {K.}~\bibnamefont {Xhani}},\ and\ \bibinfo {author}
  {\bibfnamefont {A.}~\bibnamefont {Yakimenko}},\ }\bibfield  {title} {\bibinfo
  {title} {Roadmap on atomtronics: State of the art and perspective},\ }\href
  {https://doi.org/https://doi.org/10.1116/5.0026178} {\bibfield  {journal}
  {\bibinfo  {journal} {AVS Quantum Science}\ }\textbf {\bibinfo {volume}
  {3}},\ \bibinfo {pages} {039201} (\bibinfo {year} {2021})}\BibitemShut
  {NoStop}%
\bibitem [{\citenamefont {Gauthier}\ \emph {et~al.}(2016)\citenamefont
  {Gauthier}, \citenamefont {Lenton}, \citenamefont {Parry}, \citenamefont
  {Baker}, \citenamefont {Davis}, \citenamefont {Rubinsztein-Dunlop},\ and\
  \citenamefont {Neely}}]{Gauthier2016}%
  \BibitemOpen
  \bibfield  {author} {\bibinfo {author} {\bibfnamefont {G.}~\bibnamefont
  {Gauthier}}, \bibinfo {author} {\bibfnamefont {I.}~\bibnamefont {Lenton}},
  \bibinfo {author} {\bibfnamefont {N.~M.}\ \bibnamefont {Parry}}, \bibinfo
  {author} {\bibfnamefont {M.}~\bibnamefont {Baker}}, \bibinfo {author}
  {\bibfnamefont {M.~J.}\ \bibnamefont {Davis}}, \bibinfo {author}
  {\bibfnamefont {H.}~\bibnamefont {Rubinsztein-Dunlop}},\ and\ \bibinfo
  {author} {\bibfnamefont {T.~W.}\ \bibnamefont {Neely}},\ }\bibfield  {title}
  {\bibinfo {title} {Direct imaging of a digital-micromirror device for
  configurable microscopic optical potentials},\ }\href
  {https://doi.org/https://doi.org/10.1364/OPTICA.3.001136} {\bibfield
  {journal} {\bibinfo  {journal} {Optica}\ }\textbf {\bibinfo {volume} {3}},\
  \bibinfo {pages} {1136} (\bibinfo {year} {2016})}\BibitemShut {NoStop}%
\bibitem [{\citenamefont {Gauthier}\ \emph {et~al.}(2019)\citenamefont
  {Gauthier}, \citenamefont {Szigeti}, \citenamefont {Reeves}, \citenamefont
  {Baker}, \citenamefont {Bell}, \citenamefont {Rubinsztein-Dunlop},
  \citenamefont {Davis},\ and\ \citenamefont {Neely}}]{Gauthier2019}%
  \BibitemOpen
  \bibfield  {author} {\bibinfo {author} {\bibfnamefont {G.}~\bibnamefont
  {Gauthier}}, \bibinfo {author} {\bibfnamefont {S.~S.}\ \bibnamefont
  {Szigeti}}, \bibinfo {author} {\bibfnamefont {M.~T.}\ \bibnamefont {Reeves}},
  \bibinfo {author} {\bibfnamefont {M.}~\bibnamefont {Baker}}, \bibinfo
  {author} {\bibfnamefont {T.~A.}\ \bibnamefont {Bell}}, \bibinfo {author}
  {\bibfnamefont {H.}~\bibnamefont {Rubinsztein-Dunlop}}, \bibinfo {author}
  {\bibfnamefont {M.~J.}\ \bibnamefont {Davis}},\ and\ \bibinfo {author}
  {\bibfnamefont {T.~W.}\ \bibnamefont {Neely}},\ }\bibfield  {title} {\bibinfo
  {title} {Quantitative acoustic models for superfluid circuits},\ }\href
  {https://doi.org/https://doi.org/10.1103/PhysRevLett.123.260402} {\bibfield
  {journal} {\bibinfo  {journal} {Physical Review Letters}\ }\textbf {\bibinfo
  {volume} {123}},\ \bibinfo {pages} {260402} (\bibinfo {year}
  {2019})}\BibitemShut {NoStop}%
\bibitem [{\citenamefont {Häusler}\ \emph {et~al.}(2017)\citenamefont
  {Häusler}, \citenamefont {Nakajima}, \citenamefont {Lebrat}, \citenamefont
  {Husmann}, \citenamefont {Krinner}, \citenamefont {Esslinger},\ and\
  \citenamefont {Brantut}}]{Hausler2017}%
  \BibitemOpen
  \bibfield  {author} {\bibinfo {author} {\bibfnamefont {S.}~\bibnamefont
  {Häusler}}, \bibinfo {author} {\bibfnamefont {S.}~\bibnamefont {Nakajima}},
  \bibinfo {author} {\bibfnamefont {M.}~\bibnamefont {Lebrat}}, \bibinfo
  {author} {\bibfnamefont {D.}~\bibnamefont {Husmann}}, \bibinfo {author}
  {\bibfnamefont {S.}~\bibnamefont {Krinner}}, \bibinfo {author} {\bibfnamefont
  {T.}~\bibnamefont {Esslinger}},\ and\ \bibinfo {author} {\bibfnamefont
  {J.~P.}\ \bibnamefont {Brantut}},\ }\bibfield  {title} {\bibinfo {title}
  {Scanning gate microscope for cold atomic gases},\ }\href
  {https://doi.org/https://doi.org/10.1103/physrevlett.119.030403} {\bibfield
  {journal} {\bibinfo  {journal} {Physical Review Letters}\ }\textbf {\bibinfo
  {volume} {119}},\ \bibinfo {pages} {030403} (\bibinfo {year}
  {2017})}\BibitemShut {NoStop}%
\bibitem [{\citenamefont {Deng}\ \emph {et~al.}(2022)\citenamefont {Deng},
  \citenamefont {Zhao}, \citenamefont {Liang}, \citenamefont {Chen},
  \citenamefont {Zhang},\ and\ \citenamefont {Duan}}]{Deng2022}%
  \BibitemOpen
  \bibfield  {author} {\bibinfo {author} {\bibfnamefont {M.-J.}\ \bibnamefont
  {Deng}}, \bibinfo {author} {\bibfnamefont {Y.-Y.}\ \bibnamefont {Zhao}},
  \bibinfo {author} {\bibfnamefont {Z.-X.}\ \bibnamefont {Liang}}, \bibinfo
  {author} {\bibfnamefont {J.-T.}\ \bibnamefont {Chen}}, \bibinfo {author}
  {\bibfnamefont {Y.}~\bibnamefont {Zhang}},\ and\ \bibinfo {author}
  {\bibfnamefont {X.-M.}\ \bibnamefont {Duan}},\ }\bibfield  {title} {\bibinfo
  {title} {{Maximizing energy utilization in DMD-based projection
  lithography}},\ }\href {https://doi.org/https://doi.org/10.1364/OE.448231}
  {\bibfield  {journal} {\bibinfo  {journal} {Optics Express}\ }\textbf
  {\bibinfo {volume} {30}},\ \bibinfo {pages} {4692} (\bibinfo {year}
  {2022})}\BibitemShut {NoStop}%
\bibitem [{\citenamefont {Zupancic}\ \emph {et~al.}(2016)\citenamefont
  {Zupancic}, \citenamefont {Preiss}, \citenamefont {Ruichao~Ma}, \citenamefont
  {Tai}, \citenamefont {Rispoli}, \citenamefont {Islam},\ and\ \citenamefont
  {Greiner}}]{Zupancic2016}%
  \BibitemOpen
  \bibfield  {author} {\bibinfo {author} {\bibfnamefont {P.}~\bibnamefont
  {Zupancic}}, \bibinfo {author} {\bibfnamefont {P.~M.}\ \bibnamefont
  {Preiss}}, \bibinfo {author} {\bibfnamefont {A.~L.}\ \bibnamefont
  {Ruichao~Ma}}, \bibinfo {author} {\bibfnamefont {M.~E.}\ \bibnamefont {Tai}},
  \bibinfo {author} {\bibfnamefont {M.}~\bibnamefont {Rispoli}}, \bibinfo
  {author} {\bibfnamefont {R.}~\bibnamefont {Islam}},\ and\ \bibinfo {author}
  {\bibfnamefont {M.}~\bibnamefont {Greiner}},\ }\bibfield  {title} {\bibinfo
  {title} {Ultra-precise holographic beam shaping for microscopic quantum
  control},\ }\href {https://doi.org/https://doi.org/10.1364/OE.24.013881}
  {\bibfield  {journal} {\bibinfo  {journal} {Optics Express}\ }\textbf
  {\bibinfo {volume} {24}},\ \bibinfo {pages} {13881} (\bibinfo {year}
  {2016})}\BibitemShut {NoStop}%
\bibitem [{\citenamefont {Gaunt}\ and\ \citenamefont
  {Hadzibabic}(2012)}]{Gaunt2012}%
  \BibitemOpen
  \bibfield  {author} {\bibinfo {author} {\bibfnamefont {A.~L.}\ \bibnamefont
  {Gaunt}}\ and\ \bibinfo {author} {\bibfnamefont {Z.}~\bibnamefont
  {Hadzibabic}},\ }\bibfield  {title} {\bibinfo {title} {Robust digital
  holography for ultracold atom trapping},\ }\href
  {https://doi.org/https://doi.org/10.1038/srep00721} {\bibfield  {journal}
  {\bibinfo  {journal} {Scientific Reports}\ }\textbf {\bibinfo {volume} {2}},\
  \bibinfo {pages} {721} (\bibinfo {year} {2012})}\BibitemShut {NoStop}%
\bibitem [{\citenamefont {Harte}\ \emph {et~al.}(2014)\citenamefont {Harte},
  \citenamefont {Bruce}, \citenamefont {Keeling},\ and\ \citenamefont
  {Cassettari}}]{Harte2014}%
  \BibitemOpen
  \bibfield  {author} {\bibinfo {author} {\bibfnamefont {T.}~\bibnamefont
  {Harte}}, \bibinfo {author} {\bibfnamefont {G.~D.}\ \bibnamefont {Bruce}},
  \bibinfo {author} {\bibfnamefont {J.}~\bibnamefont {Keeling}},\ and\ \bibinfo
  {author} {\bibfnamefont {D.}~\bibnamefont {Cassettari}},\ }\bibfield  {title}
  {\bibinfo {title} {Conjugate gradient minimisation approach to generating
  holographic traps for ultracold atoms},\ }\href
  {https://doi.org/https://doi.org/10.1364/OE.22.026548} {\bibfield  {journal}
  {\bibinfo  {journal} {Optics Express}\ }\textbf {\bibinfo {volume} {22}},\
  \bibinfo {pages} {26548} (\bibinfo {year} {2014})}\BibitemShut {NoStop}%
\bibitem [{\citenamefont {Pasienski}\ and\ \citenamefont
  {DeMarco}(2008)}]{DeMarco2008}%
  \BibitemOpen
  \bibfield  {author} {\bibinfo {author} {\bibfnamefont {M.}~\bibnamefont
  {Pasienski}}\ and\ \bibinfo {author} {\bibfnamefont {B.}~\bibnamefont
  {DeMarco}},\ }\bibfield  {title} {\bibinfo {title} {A high-accuracy algorithm
  for designing arbitrary holographic atom traps},\ }\href
  {https://doi.org/https://doi.org/10.1364/OE.16.002176} {\bibfield  {journal}
  {\bibinfo  {journal} {Optics Express}\ }\textbf {\bibinfo {volume} {16}},\
  \bibinfo {pages} {2176} (\bibinfo {year} {2008})}\BibitemShut {NoStop}%
\bibitem [{\citenamefont {Bowman}\ \emph {et~al.}(2017)\citenamefont {Bowman},
  \citenamefont {Harte}, \citenamefont {Chardonnet}, \citenamefont {Groot},
  \citenamefont {Denny}, \citenamefont {Goc}, \citenamefont {Anderson},
  \citenamefont {Ireland}, \citenamefont {Cassettari},\ and\ \citenamefont
  {Bruce}}]{Bowman2017}%
  \BibitemOpen
  \bibfield  {author} {\bibinfo {author} {\bibfnamefont {D.}~\bibnamefont
  {Bowman}}, \bibinfo {author} {\bibfnamefont {T.~L.}\ \bibnamefont {Harte}},
  \bibinfo {author} {\bibfnamefont {V.}~\bibnamefont {Chardonnet}}, \bibinfo
  {author} {\bibfnamefont {C.~D.}\ \bibnamefont {Groot}}, \bibinfo {author}
  {\bibfnamefont {S.~J.}\ \bibnamefont {Denny}}, \bibinfo {author}
  {\bibfnamefont {G.~L.}\ \bibnamefont {Goc}}, \bibinfo {author} {\bibfnamefont
  {M.}~\bibnamefont {Anderson}}, \bibinfo {author} {\bibfnamefont
  {P.}~\bibnamefont {Ireland}}, \bibinfo {author} {\bibfnamefont
  {D.}~\bibnamefont {Cassettari}},\ and\ \bibinfo {author} {\bibfnamefont
  {G.~D.}\ \bibnamefont {Bruce}},\ }\bibfield  {title} {\bibinfo {title}
  {High-fidelity phase and amplitude control of phase-only computer generated
  holograms using conjugate gradient minimisation},\ }\href
  {https://doi.org/https://doi.org/10.1364/OE.25.011692} {\bibfield  {journal}
  {\bibinfo  {journal} {Optics Express}\ }\textbf {\bibinfo {volume} {25}},\
  \bibinfo {pages} {11692} (\bibinfo {year} {2017})}\BibitemShut {NoStop}%
\bibitem [{\citenamefont {Ronzitti}\ \emph {et~al.}(2012)\citenamefont
  {Ronzitti}, \citenamefont {Guillon}, \citenamefont {de~Sars},\ and\
  \citenamefont {Emiliani}}]{Ronzitti2012}%
  \BibitemOpen
  \bibfield  {author} {\bibinfo {author} {\bibfnamefont {E.}~\bibnamefont
  {Ronzitti}}, \bibinfo {author} {\bibfnamefont {M.}~\bibnamefont {Guillon}},
  \bibinfo {author} {\bibfnamefont {V.}~\bibnamefont {de~Sars}},\ and\ \bibinfo
  {author} {\bibfnamefont {V.}~\bibnamefont {Emiliani}},\ }\bibfield  {title}
  {\bibinfo {title} {{LCoS nematic SLM characterization and modeling for
  diffraction efficiency optimization, zero and ghost orders suppression}},\
  }\href {https://doi.org/https://doi.org/10.1364/OE.20.017843} {\bibfield
  {journal} {\bibinfo  {journal} {Optics Express}\ }\textbf {\bibinfo {volume}
  {20}},\ \bibinfo {pages} {17843} (\bibinfo {year} {2012})}\BibitemShut
  {NoStop}%
\bibitem [{\citenamefont {Wright}\ \emph {et~al.}(2013)\citenamefont {Wright},
  \citenamefont {Blakestad}, \citenamefont {Lobb}, \citenamefont {Phillips},\
  and\ \citenamefont {Campbell}}]{Wright2013}%
  \BibitemOpen
  \bibfield  {author} {\bibinfo {author} {\bibfnamefont {K.~C.}\ \bibnamefont
  {Wright}}, \bibinfo {author} {\bibfnamefont {R.~B.}\ \bibnamefont
  {Blakestad}}, \bibinfo {author} {\bibfnamefont {C.~J.}\ \bibnamefont {Lobb}},
  \bibinfo {author} {\bibfnamefont {W.~D.}\ \bibnamefont {Phillips}},\ and\
  \bibinfo {author} {\bibfnamefont {G.~K.}\ \bibnamefont {Campbell}},\
  }\bibfield  {title} {\bibinfo {title} {Driving phase slips in a superfluid
  atom circuit with a rotating weak link},\ }\href
  {https://doi.org/http://dx.doi.org/10.1103/PhysRevLett.110.025302} {\bibfield
   {journal} {\bibinfo  {journal} {Physical Review Letters}\ }\textbf {\bibinfo
  {volume} {110}},\ \bibinfo {pages} {025302} (\bibinfo {year}
  {2013})}\BibitemShut {NoStop}%
\bibitem [{\citenamefont {Bruce}\ \emph {et~al.}(2011)\citenamefont {Bruce},
  \citenamefont {Mayoh}, \citenamefont {Smirne}, \citenamefont
  {Torralbo-Campo},\ and\ \citenamefont {Cassettari}}]{Bruce2011}%
  \BibitemOpen
  \bibfield  {author} {\bibinfo {author} {\bibfnamefont {G.~D.}\ \bibnamefont
  {Bruce}}, \bibinfo {author} {\bibfnamefont {J.}~\bibnamefont {Mayoh}},
  \bibinfo {author} {\bibfnamefont {G.}~\bibnamefont {Smirne}}, \bibinfo
  {author} {\bibfnamefont {L.}~\bibnamefont {Torralbo-Campo}},\ and\ \bibinfo
  {author} {\bibfnamefont {D.}~\bibnamefont {Cassettari}},\ }\bibfield  {title}
  {\bibinfo {title} {A smooth, holographically generated ring trap for the
  investigation of superfluidity in ultracold atoms},\ }\href
  {https://doi.org/https://dx.doi.org/10.1088/0031-8949/2011/T143/014008}
  {\bibfield  {journal} {\bibinfo  {journal} {Physica Scripta}\ }\textbf
  {\bibinfo {volume} {2011}},\ \bibinfo {pages} {014008} (\bibinfo {year}
  {2011})}\BibitemShut {NoStop}%
\bibitem [{\citenamefont {Bruce}\ \emph {et~al.}(2015)\citenamefont {Bruce},
  \citenamefont {Johnson}, \citenamefont {Cormack}, \citenamefont {Richards},
  \citenamefont {Mayoh},\ and\ \citenamefont {Cassettari}}]{Bruce2015}%
  \BibitemOpen
  \bibfield  {author} {\bibinfo {author} {\bibfnamefont {G.~D.}\ \bibnamefont
  {Bruce}}, \bibinfo {author} {\bibfnamefont {M.~Y.}\ \bibnamefont {Johnson}},
  \bibinfo {author} {\bibfnamefont {E.}~\bibnamefont {Cormack}}, \bibinfo
  {author} {\bibfnamefont {D.~A.}\ \bibnamefont {Richards}}, \bibinfo {author}
  {\bibfnamefont {J.}~\bibnamefont {Mayoh}},\ and\ \bibinfo {author}
  {\bibfnamefont {D.}~\bibnamefont {Cassettari}},\ }\bibfield  {title}
  {\bibinfo {title} {Feedback-enhanced algorithm for aberration correction of
  holographic atom traps},\ }\href
  {https://doi.org/https://dx.doi.org/10.1088/0953-4075/48/11/115303}
  {\bibfield  {journal} {\bibinfo  {journal} {Journal of Physics B: Atomic,
  Molecular and Optical Physics}\ }\textbf {\bibinfo {volume} {48}},\ \bibinfo
  {pages} {115303} (\bibinfo {year} {2015})}\BibitemShut {NoStop}%
\bibitem [{\citenamefont {Bowman}(2018)}]{Bowman2018}%
  \BibitemOpen
  \bibfield  {author} {\bibinfo {author} {\bibfnamefont {D.}~\bibnamefont
  {Bowman}},\ }\emph {\bibinfo {title} {Ultracold Atoms in Flexible Holographic
  Traps}},\ \href {http://hdl.handle.net/10023/16293} {Ph.D. thesis},\ \bibinfo
   {school} {University of St. Andrews} (\bibinfo {year} {2018})\BibitemShut
  {NoStop}%
\bibitem [{\citenamefont {Peng}\ \emph {et~al.}(2020)\citenamefont {Peng},
  \citenamefont {Choi}, \citenamefont {Padmanaban},\ and\ \citenamefont
  {Wetzstein}}]{Peng2020}%
  \BibitemOpen
  \bibfield  {author} {\bibinfo {author} {\bibfnamefont {Y.}~\bibnamefont
  {Peng}}, \bibinfo {author} {\bibfnamefont {S.}~\bibnamefont {Choi}}, \bibinfo
  {author} {\bibfnamefont {N.}~\bibnamefont {Padmanaban}},\ and\ \bibinfo
  {author} {\bibfnamefont {G.}~\bibnamefont {Wetzstein}},\ }\bibfield  {title}
  {\bibinfo {title} {Neural holography with camera-in-the-loop training},\
  }\href {https://doi.org/https://doi.org/10.1145/3414685.3417802} {\bibfield
  {journal} {\bibinfo  {journal} {ACM Trans. Graph.}\ }\textbf {\bibinfo
  {volume} {39}},\ \bibinfo {pages} {1} (\bibinfo {year} {2020})}\BibitemShut
  {NoStop}%
\bibitem [{\citenamefont {Goodman}(2017)}]{Goodman2017}%
  \BibitemOpen
  \bibfield  {author} {\bibinfo {author} {\bibfnamefont {J.~W.}\ \bibnamefont
  {Goodman}},\ }\href@noop {} {\emph {\bibinfo {title} {{Introduction to
  Fourier optics}}}}\ (\bibinfo  {publisher} {Macmillan Learning},\ \bibinfo
  {year} {2017})\BibitemShut {NoStop}%
\bibitem [{\citenamefont {Gerchberg}\ and\ \citenamefont
  {Saxton}(1972)}]{Gerchberg1972}%
  \BibitemOpen
  \bibfield  {author} {\bibinfo {author} {\bibfnamefont {R.~W.}\ \bibnamefont
  {Gerchberg}}\ and\ \bibinfo {author} {\bibfnamefont {W.~O.}\ \bibnamefont
  {Saxton}},\ }\bibfield  {title} {\bibinfo {title} {Practical algorithm for
  the determination of phase from image and diffraction plane pictures},\
  }\href@noop {} {\bibfield  {journal} {\bibinfo  {journal} {Optik}\ }\textbf
  {\bibinfo {volume} {35}},\ \bibinfo {pages} {237} (\bibinfo {year}
  {1972})}\BibitemShut {NoStop}%
\bibitem [{\citenamefont {Kim}\ \emph {et~al.}(2019)\citenamefont {Kim},
  \citenamefont {Keesling}, \citenamefont {Omran}, \citenamefont {Levine},
  \citenamefont {Bernien}, \citenamefont {Greiner}, \citenamefont {Lukin},\
  and\ \citenamefont {Englund}}]{Kim2019}%
  \BibitemOpen
  \bibfield  {author} {\bibinfo {author} {\bibfnamefont {D.}~\bibnamefont
  {Kim}}, \bibinfo {author} {\bibfnamefont {A.}~\bibnamefont {Keesling}},
  \bibinfo {author} {\bibfnamefont {A.}~\bibnamefont {Omran}}, \bibinfo
  {author} {\bibfnamefont {H.}~\bibnamefont {Levine}}, \bibinfo {author}
  {\bibfnamefont {H.}~\bibnamefont {Bernien}}, \bibinfo {author} {\bibfnamefont
  {M.}~\bibnamefont {Greiner}}, \bibinfo {author} {\bibfnamefont {M.~D.}\
  \bibnamefont {Lukin}},\ and\ \bibinfo {author} {\bibfnamefont {D.~R.}\
  \bibnamefont {Englund}},\ }\bibfield  {title} {\bibinfo {title} {Large-scale
  uniform optical focus array generation with a phase spatial light
  modulator},\ }\href {https://doi.org/https://doi.org/10.1364/OL.44.003178}
  {\bibfield  {journal} {\bibinfo  {journal} {Optics Letters}\ }\textbf
  {\bibinfo {volume} {44}},\ \bibinfo {pages} {3178} (\bibinfo {year}
  {2019})}\BibitemShut {NoStop}%
\bibitem [{\citenamefont {Milewski}\ \emph {et~al.}(2007)\citenamefont
  {Milewski}, \citenamefont {Engström},\ and\ \citenamefont
  {Bengtsson}}]{Bengtsson2007}%
  \BibitemOpen
  \bibfield  {author} {\bibinfo {author} {\bibfnamefont {G.}~\bibnamefont
  {Milewski}}, \bibinfo {author} {\bibfnamefont {D.}~\bibnamefont
  {Engström}},\ and\ \bibinfo {author} {\bibfnamefont {J.}~\bibnamefont
  {Bengtsson}},\ }\bibfield  {title} {\bibinfo {title} {Diffractive optical
  elements designed for highly precise far-field generation in the presence of
  artifacts typical for pixelated spatial light modulators},\ }\href
  {https://doi.org/https://doi.org/10.1364/AO.46.000095} {\bibfield  {journal}
  {\bibinfo  {journal} {Applied Optics}\ }\textbf {\bibinfo {volume} {46}},\
  \bibinfo {pages} {95} (\bibinfo {year} {2007})}\BibitemShut {NoStop}%
\bibitem [{\citenamefont {Persson}\ \emph {et~al.}(2012)\citenamefont
  {Persson}, \citenamefont {Engström},\ and\ \citenamefont
  {Goksör}}]{Persson2012}%
  \BibitemOpen
  \bibfield  {author} {\bibinfo {author} {\bibfnamefont {M.}~\bibnamefont
  {Persson}}, \bibinfo {author} {\bibfnamefont {D.}~\bibnamefont {Engström}},\
  and\ \bibinfo {author} {\bibfnamefont {M.}~\bibnamefont {Goksör}},\
  }\bibfield  {title} {\bibinfo {title} {Reducing the effect of pixel crosstalk
  in phase only spatial light modulators},\ }\href
  {https://doi.org/https://doi.org/10.1364/OE.20.022334} {\bibfield  {journal}
  {\bibinfo  {journal} {Optics Express}\ }\textbf {\bibinfo {volume} {20}},\
  \bibinfo {pages} {22334} (\bibinfo {year} {2012})}\BibitemShut {NoStop}%
\bibitem [{\citenamefont {Moser}\ \emph {et~al.}(2019)\citenamefont {Moser},
  \citenamefont {Ritsch-Marte},\ and\ \citenamefont {Thalhammer}}]{Moser2019}%
  \BibitemOpen
  \bibfield  {author} {\bibinfo {author} {\bibfnamefont {S.}~\bibnamefont
  {Moser}}, \bibinfo {author} {\bibfnamefont {M.}~\bibnamefont
  {Ritsch-Marte}},\ and\ \bibinfo {author} {\bibfnamefont {G.}~\bibnamefont
  {Thalhammer}},\ }\bibfield  {title} {\bibinfo {title} {Model-based
  compensation of pixel crosstalk in liquid crystal spatial light modulators},\
  }\href {https://doi.org/https://doi.org/10.1364/OE.27.025046} {\bibfield
  {journal} {\bibinfo  {journal} {Optics Express}\ }\textbf {\bibinfo {volume}
  {27}},\ \bibinfo {pages} {25046} (\bibinfo {year} {2019})}\BibitemShut
  {NoStop}%
\bibitem [{\citenamefont {Pushkina}\ \emph {et~al.}(2020)\citenamefont
  {Pushkina}, \citenamefont {Costa-Filho}, \citenamefont {Maltese},\ and\
  \citenamefont {Lvovsky}}]{Pushkina2020}%
  \BibitemOpen
  \bibfield  {author} {\bibinfo {author} {\bibfnamefont {A.~A.}\ \bibnamefont
  {Pushkina}}, \bibinfo {author} {\bibfnamefont {J.~I.}\ \bibnamefont
  {Costa-Filho}}, \bibinfo {author} {\bibfnamefont {G.}~\bibnamefont
  {Maltese}},\ and\ \bibinfo {author} {\bibfnamefont {A.~I.}\ \bibnamefont
  {Lvovsky}},\ }\bibfield  {title} {\bibinfo {title} {Comprehensive model and
  performance optimization of phase-only spatial light modulators},\ }\href
  {https://doi.org/https://dx.doi.org/10.1088/1361-6501/aba56b} {\bibfield
  {journal} {\bibinfo  {journal} {Measurement Science and Technology}\ }\textbf
  {\bibinfo {volume} {31}},\ \bibinfo {pages} {125202} (\bibinfo {year}
  {2020})}\BibitemShut {NoStop}%
\bibitem [{\citenamefont {Guesmi}\ and\ \citenamefont
  {Žídek}(2021)}]{Guesmi2021}%
  \BibitemOpen
  \bibfield  {author} {\bibinfo {author} {\bibfnamefont {M.}~\bibnamefont
  {Guesmi}}\ and\ \bibinfo {author} {\bibfnamefont {K.}~\bibnamefont
  {Žídek}},\ }\bibfield  {title} {\bibinfo {title} {Calibration of the pixel
  crosstalk in spatial light modulators for 4f pulse shaping},\ }\href
  {https://doi.org/https://doi.org/10.1364/AO.434309} {\bibfield  {journal}
  {\bibinfo  {journal} {Applied Optics}\ }\textbf {\bibinfo {volume} {60}},\
  \bibinfo {pages} {7648} (\bibinfo {year} {2021})}\BibitemShut {NoStop}%
\bibitem [{\citenamefont {Moreno}\ \emph {et~al.}(2021)\citenamefont {Moreno},
  \citenamefont {Sánchez-López}, \citenamefont {Davis},\ and\ \citenamefont
  {Cottrell}}]{Moreno2021}%
  \BibitemOpen
  \bibfield  {author} {\bibinfo {author} {\bibfnamefont {I.}~\bibnamefont
  {Moreno}}, \bibinfo {author} {\bibfnamefont {M.~D.~M.}\ \bibnamefont
  {Sánchez-López}}, \bibinfo {author} {\bibfnamefont {J.~A.}\ \bibnamefont
  {Davis}},\ and\ \bibinfo {author} {\bibfnamefont {D.~M.}\ \bibnamefont
  {Cottrell}},\ }\bibfield  {title} {\bibinfo {title} {Simple method to
  evaluate the pixel crosstalk caused by fringing field effect in
  liquid-crystal spatial light modulators},\ }\href
  {https://doi.org/https://doi.org/10.1186/s41476-021-00174-7} {\bibfield
  {journal} {\bibinfo  {journal} {Journal of the European Optical Society-Rapid
  Publications}\ }\textbf {\bibinfo {volume} {17}},\ \bibinfo {pages} {27}
  (\bibinfo {year} {2021})}\BibitemShut {NoStop}%
\bibitem [{\citenamefont {van Bijnen}(2013)}]{VanBijnen2013}%
  \BibitemOpen
  \bibfield  {author} {\bibinfo {author} {\bibfnamefont {R.}~\bibnamefont {van
  Bijnen}},\ }\emph {\bibinfo {title} {Quantum engineering with ultracold
  atoms}},\ \href {https://doi.org/https://doi.org/10.6100/IR754785} {Ph.D.
  thesis},\ \bibinfo  {school} {Technische Universiteit Eindhoven} (\bibinfo
  {year} {2013})\BibitemShut {NoStop}%
\bibitem [{\citenamefont {Senthilkumaran}(2018)}]{Sent2018}%
  \BibitemOpen
  \bibfield  {author} {\bibinfo {author} {\bibfnamefont {P.}~\bibnamefont
  {Senthilkumaran}},\ }\href
  {https://doi.org/https://dx.doi.org/10.1088/978-0-7503-1698-9} {\emph
  {\bibinfo {title} {Vortices in computational optics}}}\ (\bibinfo
  {publisher} {IOP Publishing},\ \bibinfo {year} {2018})\BibitemShut {NoStop}%
\bibitem [{\citenamefont {Clark}\ \emph {et~al.}(2016)\citenamefont {Clark},
  \citenamefont {Offer}, \citenamefont {Franke-Arnold}, \citenamefont
  {Arnold},\ and\ \citenamefont {Radwell}}]{Clark2016}%
  \BibitemOpen
  \bibfield  {author} {\bibinfo {author} {\bibfnamefont {T.~W.}\ \bibnamefont
  {Clark}}, \bibinfo {author} {\bibfnamefont {R.~F.}\ \bibnamefont {Offer}},
  \bibinfo {author} {\bibfnamefont {S.}~\bibnamefont {Franke-Arnold}}, \bibinfo
  {author} {\bibfnamefont {A.~S.}\ \bibnamefont {Arnold}},\ and\ \bibinfo
  {author} {\bibfnamefont {N.}~\bibnamefont {Radwell}},\ }\bibfield  {title}
  {\bibinfo {title} {Comparison of beam generation techniques using a phase
  only spatial light modulator},\ }\href
  {https://doi.org/https://doi.org/10.1364/OE.24.006249} {\bibfield  {journal}
  {\bibinfo  {journal} {Optics Express}\ }\textbf {\bibinfo {volume} {24}},\
  \bibinfo {pages} {6249} (\bibinfo {year} {2016})}\BibitemShut {NoStop}%
\bibitem [{\citenamefont {Polak}\ and\ \citenamefont
  {Ribiere}(1969)}]{polak1969}%
  \BibitemOpen
  \bibfield  {author} {\bibinfo {author} {\bibfnamefont {E.}~\bibnamefont
  {Polak}}\ and\ \bibinfo {author} {\bibfnamefont {G.}~\bibnamefont
  {Ribiere}},\ }\bibfield  {title} {\bibinfo {title} {Note sur la convergence
  de m\'{e}thodes de directions conjugu\'{e}es},\ }\href
  {https://doi.org/https://doi.org/10.1051/M2AN%2F196903R100351} {\bibfield
  {journal} {\bibinfo  {journal} {Rev. Fr. Inform. Rech. Op\'{e}r.}\ }\textbf
  {\bibinfo {volume} {3}},\ \bibinfo {pages} {35} (\bibinfo {year}
  {1969})}\BibitemShut {NoStop}%
\bibitem [{\citenamefont {Geiger}\ \emph {et~al.}(2012)\citenamefont {Geiger},
  \citenamefont {Moosmann}, \citenamefont {Ömer Car},\ and\ \citenamefont
  {Schuster}}]{Geiger2012}%
  \BibitemOpen
  \bibfield  {author} {\bibinfo {author} {\bibfnamefont {A.}~\bibnamefont
  {Geiger}}, \bibinfo {author} {\bibfnamefont {F.}~\bibnamefont {Moosmann}},
  \bibinfo {author} {\bibnamefont {Ömer Car}},\ and\ \bibinfo {author}
  {\bibfnamefont {B.}~\bibnamefont {Schuster}},\ }\bibfield  {title} {\bibinfo
  {title} {Automatic camera and range sensor calibration using a single shot},\
  }in\ \href {https://doi.org/https://doi.org/10.1109/ICRA.2012.6224570} {\emph
  {\bibinfo {booktitle} {2012 IEEE International Conference on Robotics and
  Automation}}}\ (\bibinfo {year} {2012})\ pp.\ \bibinfo {pages}
  {3936--3943}\BibitemShut {NoStop}%
\bibitem [{\citenamefont {Senthilkumaran}\ \emph {et~al.}(2005)\citenamefont
  {Senthilkumaran}, \citenamefont {Wyrowski},\ and\ \citenamefont
  {Schimmel}}]{Schimmel2005}%
  \BibitemOpen
  \bibfield  {author} {\bibinfo {author} {\bibfnamefont {P.}~\bibnamefont
  {Senthilkumaran}}, \bibinfo {author} {\bibfnamefont {F.}~\bibnamefont
  {Wyrowski}},\ and\ \bibinfo {author} {\bibfnamefont {H.}~\bibnamefont
  {Schimmel}},\ }\bibfield  {title} {\bibinfo {title} {{Vortex stagnation
  problem in iterative Fourier transform algorithms}},\ }\href
  {https://doi.org/https://doi.org/10.1016/j.optlaseng.2004.06.002} {\bibfield
  {journal} {\bibinfo  {journal} {Optics and Lasers in Engineering}\ }\textbf
  {\bibinfo {volume} {43}},\ \bibinfo {pages} {43} (\bibinfo {year}
  {2005})}\BibitemShut {NoStop}%
\bibitem [{\citenamefont {Lanczos}(1938)}]{Lanczos1938}%
  \BibitemOpen
  \bibfield  {author} {\bibinfo {author} {\bibfnamefont {C.}~\bibnamefont
  {Lanczos}},\ }\bibfield  {title} {\bibinfo {title} {Trigonometric
  interpolation of empirical and analytical functions},\ }\href
  {https://doi.org/https://doi.org/10.1002/sapm1938171123} {\bibfield
  {journal} {\bibinfo  {journal} {Journal of Mathematics and Physics}\ }\textbf
  {\bibinfo {volume} {17}},\ \bibinfo {pages} {123} (\bibinfo {year}
  {1938})}\BibitemShut {NoStop}%
\bibitem [{\citenamefont {Schott}(2022{\natexlab{a}})}]{nsf11}%
  \BibitemOpen
  \bibfield  {author} {\bibinfo {author} {\bibnamefont {Schott}},\ }\href
  {https://www.schott.com/shop/advanced-optics/en/Optical-Glass/N-SF11/c/glass-N-SF11}
  {\bibinfo {title} {{N-SF11}}},\ \bibinfo {howpublished}
  {\url{https://www.schott.com/shop/advanced-optics/en/Optical-Glass/N-SF11/c/glass-N-SF11}}
  (\bibinfo {year} {2022}{\natexlab{a}})\BibitemShut {NoStop}%
\bibitem [{\citenamefont {Schott}(2022{\natexlab{b}})}]{nsk2}%
  \BibitemOpen
  \bibfield  {author} {\bibinfo {author} {\bibnamefont {Schott}},\ }\href
  {https://www.schott.com/shop/advanced-optics/en/Optical-Glass/N-SK2/c/glass-N-SK2}
  {\bibinfo {title} {{N-SK2}}},\ \bibinfo {howpublished}
  {\url{https://www.schott.com/shop/advanced-optics/en/Optical-Glass/N-SK2/c/glass-N-SK2}}
  (\bibinfo {year} {2022}{\natexlab{b}})\BibitemShut {NoStop}%
\end{thebibliography}%

\section*{Acknowledgements}
We acknowledge support by the Engineering and Physical Sciences Research Council (EPSRC) through the Quantum Technology Hub in Quantum Computing and Simulation [grant number EP/T001062/1], the 2020-2021 Doctoral Training Partnership [EP/T517811/1],the Programme Grant DesOEQ [EP/P009565/1] and the New Investigator Grant [EP/T027789/1]. We also would like to thank Graham Bruce, Dimitrios Tsevas, Elliot Diamond-Hitchcock, Nicholas Spong, Jonathan Pritchard and Aidan Arnold for helpful discussions. 

\section*{Author contributions statement}
P.S. implemented the simulations, built the experimental setup and performed the measurements. A.L.R., E.H. and S.K. supervised the project and helped analysing the results. All authors contributed to the preparation of the  manuscript. 

\section*{Additional information}
The authors declare no competing interests.

\section*{Data availability statement} 
The datasets generated and/or analysed during the current study are not publicly available due the size (280 GB in images) but are available from the corresponding author on reasonable request.

\newpage

\renewcommand{\thefigure}{S\arabic{figure}}
\setcounter{figure}{0}

\renewcommand{\theequation}{S\arabic{equation}}
\setcounter{equation}{0}

\phantomsection

\section*{Experimental setup}
\begin{figure}[!htb]
\centering{}
\includegraphics[width=14cm]{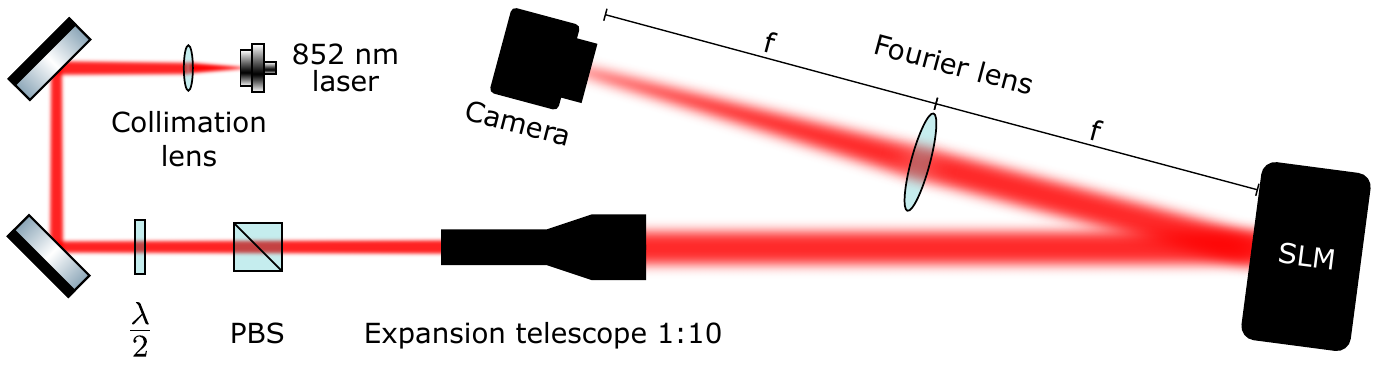}
\caption{Schematic of the experimental setup.}
\label{fig:methods_setup}
\end{figure}

Light at wavelength $\lambda=\unit[852]{nm}$ from a single-mode fibre is collimated by a triplet lens (Melles Griot 06 GLC 001) with a specified wavefront distortion of $<\frac{\lambda}{4}$ and is expanded by a telescope (Thorlabs GBE10-B) to a diameter of $\unit[ 9.4]{mm}$ at the SLM.  The light is polarised along the horizontal plane by a polarising beam splitter (see Fig. \ref{fig:methods_setup}). The beam is reflected by the SLM (Hamamatsu X13138-07, $\unit[12.5]{\mu m}$ pixel pitch, $1272 \times 1024$ pixels) at an angle of $\sim 10^{\circ}$ and is focussed onto the camera (Matrix Vision mvBlueFOX3-1012dG, $\unit[3.75]{\mu m}$ pixel pitch,$1280 \times 960$ pixels) by the Fourier lens (Thorlabs ACT508-250-B). 

\captionsetup[figure]{aboveskip=1pt}

\section{Wavefront measurement}
\label{sec:wf_meas}

\begin{figure}[!t] \centering{\phantomsubcaption\label{fig:int_scheme}\phantomsubcaption\label{fig:wf_int}}
    \includegraphics[width=14cm]{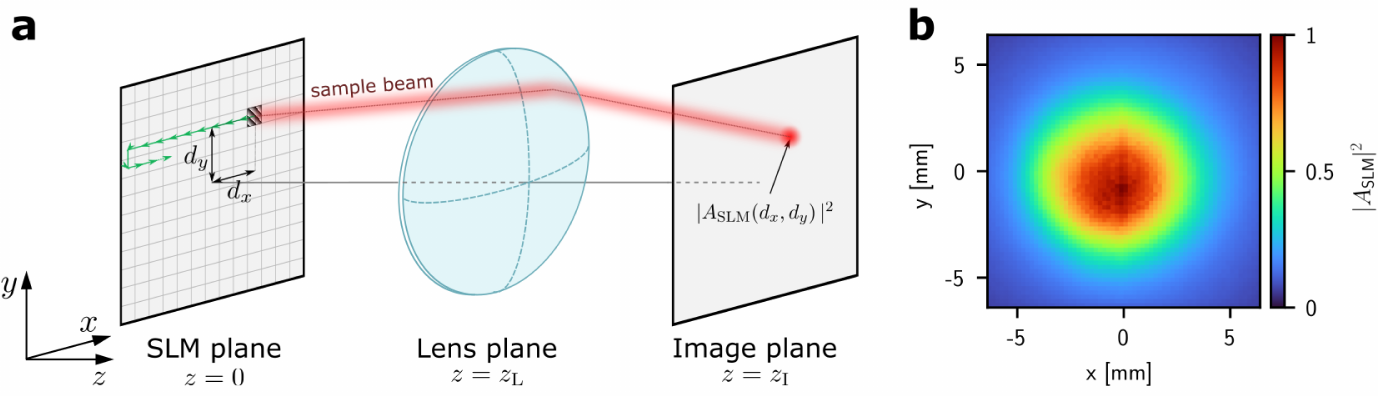}
    \caption{(\subref{fig:int_scheme}) Scheme illustrating the measurement of the laser intensity profile by displaying a series of apertures containing a linear gradient on the SLM \protect\cite{Clark2016} (Fig. adapted from Zupancic et al. \cite{Zupancic2016}). (\subref{fig:wf_int}) Resulting laser intensity profile.}
\label{fig:int}
\end{figure}

To generate experimental light potentials that match the simulated ones, it is essential to precisely know the wavefront of the light reflected by the SLM and the intensity profile of the incident laser beam. We measure the constant phase, $\varphi_{\text{C}}$, across the SLM using a scheme introduced in a previous study by Zupancic et al. \cite{Zupancic2016}. To measure the intensity profile across the SLM, we sample the local intensity by displaying a square pattern on an area of $32 \times 32$ pixels containing a linear phase gradient (see Fig. \ref{fig:int_scheme}), while on the remaining area of the SLM, a flat phase is displayed. This phase gradient generates a diffraction spot away from the optical axis, and the light incident onto the remaining area of the SLM collects on the optical axis. We vary the position of the square pattern, $d_x$ and $d_y$, across the entire area of the SLM and measure the intensity of each diffraction spot, $|A_{\text{SLM}}\!\left(d_x, d_y\right)|^2$, on the camera, and as a result, the intensity profile of the laser beam across the SLM is reconstructed (Fig. \ref{fig:wf_int}) \cite{Clark2016}. The position of the square is varied on an equally spaced grid using $64 \times 64$ measurements. The diffraction angle of the linear phase gradient is $\alpha_x=\alpha_y=0.5^{\circ}$ both in x- and y-direction. Initially, the square is displayed at the centre of the SLM and a Gaussian is fitted to the resulting diffraction spot on the camera, in a square region of interest of 300 camera pixels. The intensity of each spot is calculated as the sum of all pixel values in the region of interest.

\begin{figure}[!tb] \centering{\phantomsubcaption\label{fig:phi_scheme}\phantomsubcaption\label{fig:wf_phi}}
    \includegraphics[width=14cm]{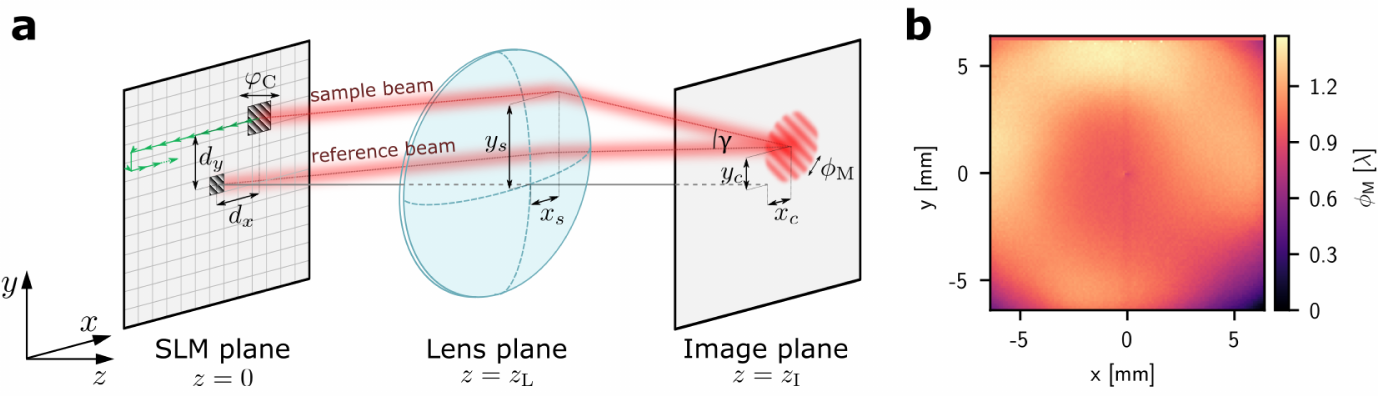}
    \caption{(\subref{fig:phi_scheme}) Scheme illustrating the measurement of the constant phase at the SLM using an interferometric approach by displaying a sequence of patterns on sub-regions of the SLM (adapted from Zupancic et al. \cite{Zupancic2016}). (\subref{fig:wf_phi}) Resulting measured phase, $\varphi_{\text{M}}$, expressed in units of $\lambda$.}
\label{fig:wfs}
\end{figure}

To measure the constant phase, the position of a square sample phase pattern is varied across the entire area of the SLM, similar to our scheme used to measure the intensity. In addition, a reference square pattern is displayed at the centre of the SLM (see Fig. \ref{fig:phi_scheme}). The beams originating from the two phase patterns interfere at the camera, causing sine-shaped fringes. The spatial phase, $\phi_{\text{M}}$, of this interference pattern is detected by fitting a 2D sine pattern to the camera image \cite{Zupancic2016}
\begin{equation}
I_{\text{IMG}}\!\left( x, y \right) = A^2 + B^2 + 2AB\cos \left[ k \left( x\sin\gamma_x + y\sin\gamma_y\right) + \phi_{\text{M}} \right],
\label{eq:sine_fit}
\end{equation}
where $\gamma_x = \arctan\left( d_x / f \right)$ and $\gamma_y = \arctan\left( d_y / f \right)$. Here, $d_x$ and $d_y$ are the position of the sample pattern with respect to the reference pattern and $f$ is the focal length of the Fourier lens. $A$ and $B$ are the amplitudes of the diffracted beams caused by the reference and the sample square pattern, respectively. Assuming perfect positioning of the lens at $z=f$ and the camera at $z=2f$ and assuming a thin and parabolic lens, the measured phase, $\phi_{\text{M}}$, corresponds to the phase difference between the reference aperture and the sampling aperture $\varphi_{\text{C}}=\phi_{\text{M}}$. The parameters $A$, $B$ and $\phi_{\text{M}}$ are fitted while $\gamma_x$ and $\gamma_y$ are calculated. Due to the Gaussian shape of the beam incident onto the SLM, the intensity of the light at the SLM drops off significantly towards the edges. This causes the intensity of the sampling beam $B$ to become very small compared to $A$ as the sampling aperture moves away from the centre of the SLM, resulting in a low contrast $2AB$ of the interference pattern and a poor fit. To counteract this, the size of the sampling patch is increased as it moves away from the centre of the SLM to keep the power contained in the sampling aperture equal to the power contained in the reference aperture. This increases the contrast of the interference pattern on the camera and improves the measurement of the phase at darker regions of the SLM. We use $124 \times 124$ measurements, equally spaced across the SLM with a reference phase pattern of $16 \times 16$ SLM pixels, resulting in the measured constant phase $\varphi_{\text{C}}$ in Fig. \ref{fig:wf_phi}.  Displaying $-\varphi_{\text{C}}$ on the SLM and re-running the measurement results in a flat phase of $\sim \lambda / 40$ RMS error. 
As the measured intensity and phase have $32 \times 32$ and $124 \times 124$ data points, they are up-scaled to the native resolution of the SLM (central $1024 \times 1024$ pixels) using fourth-order Lanczos interpolation \cite{Lanczos1938}. Before up-scaling, the phase is unwrapped and both measurements are smoothed using a $3\times 3$ uniform filter. It takes approximately 30 minutes to calibrate the intensity pattern and 2 hours to obtain the phase calibration. 

\section{Angular spectrum method}
\label{si:asm}
We implement the ASM to simulate the propagation of light in our CG minimisation. First, the electric field at the SLM plane, $E_{\text{SLM}}\!\left(x, y\right)$, is propagated to the lens plane and multiplied by the aperture, $A_{\text{L}}\!\left(x, y\right)$, and phase, $\phi_{\text{L}}\!\left(x, y\right)$, of the lens using the relation \cite{Goodman2017}
\begin{equation}
E\!\left( x, y, z_{\text{L}} \right) = \mathcal{F}^{-1}\Bigl\{\mathcal{F}\left\{ E_{\text{SLM}}\!\left(x, y\right)\right\}H\!\left(\kappa_x', \kappa_y', z_{\text{L}}\right)
\label{eq:asm1}\Bigr\}A_{\text{L}}\!\left(x, y\right)\exp\left[ i\phi_{\text{L}}\!\left(x, y\right)\right].
\end{equation}
Here, $\kappa_x'$ and $\kappa_y'$ are the spatial frequencies, $E\left( x, y, z_{\text{L}} \right)$ is the electric field in the lens plane just after the lens and $z_{\text{L}}$ is the distance between the SLM plane and the lens plane. $A_{\text{L}}\!\left(x, y\right) = circ\!\left( r\right)$ is the circular aperture of the lens with radius $r$ and $\phi_{\text{L}}\!\left(x, y\right)$ is the phase delay caused by the lens. The transfer function, $H\left(\kappa_x', \kappa_y', \Delta z\right)$, is given by \cite{Goodman2017}
\begin{equation}
H\!\left(\kappa_x', \kappa_y', \Delta z\right) =\begin{cases} \exp\left[ 2\pi i\frac{\Delta z}{\lambda}\sqrt{1-\left(\lambda \kappa_x'\right)^2-\left(\lambda \kappa_y'\right)^2}\right] & \text{if $\sqrt{\kappa_x'^2+\kappa_y'^2} < \frac{1}{\lambda}$}.\\
0 & \text{otherwise}.
\end{cases}
\label{eq:asm2}
\end{equation}
with propagation distance, $\Delta z$. The resulting electric field, $E\!\left( x, y, z_{\text{L}} \right)$, is then propagated to the image plane using \cite{Goodman2017}
\begin{equation}
E\!\left( x, y, z_{\text{I}} \right) = \mathcal{F}^{-1}\left\{\mathcal{F}\left\{ E\!\left(x, y, z_{\text{L}}\right)\right\}H\!\left(\kappa_x', \kappa_y', z_{\text{I}}-z_{\text{L}}\right)\right\},
\label{eq:asm3}
\end{equation}
where $E\!\left( x, y, z_{\text{I}} \right)$ is the resulting electric field in the image plane and $\Delta z=z_{\text{I}}-z_{\text{L}}$ is the distance between the lens plane and the image plane (see Fig. 1a).

Using the ASM instead of the Fourier transform enables us to model the lens accurately.
Specifically, we use a doublet lens with three spherical surfaces (Thorlabs ACT508-250-B) which causes a phase delay \cite{Goodman2017}
\begin{equation}
\phi_{\text{L}}\!\left(x, y\right) = \frac{2\pi}{\lambda}\left[\Delta_{12}\left(x, y\right)\left(n_1-1\right)+\Delta_{23}\left(x, y\right)\left(n_2-1\right)\right],
\label{eq:doublet}
\end{equation}
where $\Delta_{12}$ and $\Delta_{23}$ are the lens thicknesses and $n_1=1.59847$ and $n_2=1.76182$ the refractive indices of the crown and the flint glass \cite{nsf11, nsk2}, respectively. The lens thicknesses are given by \cite{Goodman2017}
\begin{equation}
\Delta_{ab}\!\left(x, y\right) = -R_{a}\!\left(1-\sqrt{1-\frac{x^2+y^2}{R_{a}^2}}\right) + R_{b}\!\left(1-\sqrt{1-\frac{x^2+y^2}{R_{b}^2}}\right)
\label{eq:delta1}
\end{equation}
with the radii of the spherical surfaces $R_1=137.7\,\text{mm}$, $R_2=-R_1$ and $R_3=-930.4\,\text{mm}$.
The phase of the doublet, $\phi_{\text{L}}\left(x, y\right)$, deviates from the idealised phase of the lens \cite{Goodman2017}
\begin{equation}
\phi_{\text{Q}}\!\left(x, y\right) = -\frac{\pi}{\lambda f}\left(x^2 + y^2\right),
\label{eq:ideal_lens}
\end{equation}
with the focal length, $f=250\,\text{mm}$, by $2.8\,\lambda$ (peak-to-valley) across the aperture of the lens (48.3 mm).

In our numerical implementation, we pad the array representing the SLM field with zeros to match the size of the SLM plane with the aperture of the lens used in our experiment.
This increases the computational complexity as the matrix size increases from $2048\times 2048$ to $3864\times3864$.
When using the FFT, the matrix representing the SLM plane of $1024\times 1024$ pixels is zero-padded to $2048\times 2048$ pixels, resulting in a pixel spacing $p_{\text{IMG}}=\frac{\lambda f}{2Np_{\text{SLM}}}=\unit[8.32]{\mu m}$ in the image plane, with the number of SLM pixels, $N$, in each dimension and SLM pixel pitch, $p_{\text{SLM}}$.
With the ASM, the pixel size in the SLM plane equals the pixel size in the image plane.
To achieve a similar spatial resolution in the output plane using the ASM, each SLM pixel of $\unit[12.5]{\mu m}$ size is sub-resolved computationally into $2\times 2$ pixels which increases the number of pixels to $7728\times 7728$.
We use a GPU (Nvidia RTX A5000 24 GB) to accelerate our calculations.

\subsection{ASM wavefront correction}
\label{sec:corr_wf_meas}
Our method to measure the constant phase, $\varphi_{\text{C}}$, requires the lens to be parabolic and assumes perfect placement of the lens and the camera.
Using equation \ref{eq:sine_fit}, the measured phase, $\phi_{\text{M}}\!\left(x, y\right)$, includes the phase difference caused by the distorted wavefront at the SLM and the phase differences caused by a non-parabolic lens and a displacement of the camera along the optical axis.
As the ASM is capable of modelling the doublet lens and a displaced camera, it is important to separate these phase differences and the wavefront at the SLM, $\phi_{\text{C}}\!\left(x, y\right)$, from each other. 

To implement the ASM, we calculate a corrective phase, $\phi_{\text{ASM}}\!\left(x, y\right)$, which only models the phase caused by the displaced, non-parabolic lens and the displaced camera, assuming a flat wavefront at the SLM.
To do so, we calculate the path length of every sample beam between the lens and a fixed point in the image plane as well as the phase delay each sample beam collects when passing through the lens, $\phi_{\text{L}}\!\bigl(x_{s}\!\left(x\right), y_{s}\!\left(y\right)\bigr)$.
\begin{equation}
\phi_{\text{ASM}}\!\left(x, y\right) = \frac{2 \pi}{\lambda}\sqrt{\bigl[z_{\text{I}}-z_{\text{L}}\bigr]^2 + \bigl[x_{s}\!\left(x\right) - x_c\bigr]^2 + \bigl[y_{s}\!\left(y\right) - y_c\bigr]^2} + \phi_{\text{L}}\!\bigl(x_{s}\!\left(x\right), y_{s}\!\left(y\right)\bigr),
\label{eq:phi_asm}
\end{equation}
with the position of the sample beam on the lens $x_{s}\!\left(x\right)=x + z_{\text{L}}\tan\!\left(\alpha_x\right)$ and $y_{s}\!\left(y\right)=y + z_{\text{L}}\tan\left(\alpha_y\right)$, where $\alpha_x$ and $\alpha_y$ are the diffraction angles of the linear phase gradient in x- and y-direction, respectively.
The phase is sampled at a point in the image plane with co-ordinates $x_c=f \tan\!\left(\alpha_x\right)$ and $y_c=f \tan\!\left(\alpha_y\right)$.
We then subtract the corrective phase pattern, $\!\phi_{\text{ASM}}\left(x, y\right)$, from the measured constant phase to obtain the wavefront at the SLM, $\phi_{\text{C}}\!\left(x, y\right)=\phi_{\text{M}}\!\left(x, y\right)-\phi_{\text{ASM}}\!\left(x, y\right)$.

\end{document}